\documentclass[aps,prl,twocolumn,superscriptaddress,nofootinbib]{revtex4-2}
\usepackage[colorlinks]{hyperref}
\usepackage{amsmath, mathtools, cleveref, comment, booktabs, xcolor,soul,amsfonts}
\usepackage{ulem}

\usepackage{float,placeins,longtable}

\usepackage{overpic}

\hypersetup{
    colorlinks=true, 
    linkcolor=blue,  
    citecolor=blue,  
    filecolor=magenta,
    urlcolor=blue    
}

\begin{document}

\title{Electromagnetic form factors and structure of the $T_{bb}$ tetraquark from lattice QCD}

\author{Ivan Vujmilovic}
\email[]{ivan.vujmilovic@ijs.si}
\affiliation{Jožef Stefan Institute, Jamova 39, 1000 Ljubljana, Slovenia}
\affiliation{Faculty of Mathematics and Physics, University of Ljubljana, 1000 Ljubljana, Slovenia}

\author{Sara Collins}
\email[]{sara.collins@ur.de}
\affiliation{Institut f\"ur Theoretische Physik, Universit\"at Regensburg, 93040 Regensburg, Germany}

\author{Luka Leskovec}
\email[]{luka.leskovec@ijs.si}
\affiliation{Jožef Stefan Institute, Jamova 39, 1000 Ljubljana, Slovenia}
\affiliation{Faculty of Mathematics and Physics, University of Ljubljana, 1000 Ljubljana, Slovenia}

\author{Sasa Prelovsek}
\email[]{sasa.prelovsek@ijs.si}
\affiliation{Jožef Stefan Institute, Jamova 39, 1000 Ljubljana, Slovenia}
\affiliation{Faculty of Mathematics and Physics, University of Ljubljana, 1000 Ljubljana, Slovenia}

\begin{abstract}
We present the first lattice QCD determination of the electromagnetic form factors of the exotic tetraquark $T_{bb} \ (bb \bar u \bar d)$ with quantum numbers $I( J^P ) = 0( 1^+ )$.  The extracted form factors encode information about its internal structure, including the charge distribution and the magnetic dipole moments, determined separately for the light and heavy quarks. Our results provide evidence in favor of it being a bound state consisting of a compact heavy diquark $[bb]$ in a color-antitriplet with spin one, and a light antidiquark $[\bar u \bar d]$ in a color-triplet with spin zero. The charge radius of $T_{bb}$ is found to be significantly smaller than the combined charge radii of $B$ and $B^*$ mesons. These two comprise the lowest-lying threshold $BB^*$ in the channel we are considering, and their electric charge form factors are also determined. The computations were performed on a single CLS ensemble with $N_f = 2+1$ dynamical quarks and a lattice spacing of approximately $a \approx0.064 \ \mathrm{fm}$ at the pion mass $m_\pi \approx 290 \ \mathrm{MeV}$.
\end{abstract}

\maketitle

\textbf{\textit{Introduction}}: Quantum Chromodynamics (QCD), as an integral part of the Standard Model of particle physics, has achieved enormous success in explaining the existence of hadrons---mesons and baryons that are bound states
or resonances. In a simple quark model picture these are built either from a single quark-antiquark pair ($\bar q_1 q_2$) or three quarks ($q_1 q_2 q_3$), respectively \cite{ref:gellmann, Zweig:1964jf}. QCD also allows for the existence of more exotic states \cite{Jaffe:2004ph, Lebed:2016hpi}, e.g. tetraquarks ($\bar q_1 \bar q_2  q_3  q_4$), pentaquarks ($q_1\bar{q}_2q_3q_4q_5$), hybrid mesons ($\bar q_1 g q_2$) or glueball-like mesons, the latter built mainly from gluons; many of these have been discovered experimentally \cite{ParticleDataGroup:2024cfk,ref:x3872,Belle:2007hrb,BESIII:2013ris,BaBar:2005hhc,LHCb:2015yax,LHCb:2019kea,CDF:2009jgo,LHCb:2016axx,LHCb:2021vvq}. Nevertheless the quest to determine how the exotic  states populate  the dense spectrum arising from QCD is one of the most active avenues of research in hadronic physics \cite{Bicudo:2022cqi,Francis:2024fwf,Brambilla:2019esw,Bulava:2022ovd}.\\ 
\indent The as-yet-experimentally-undiscovered doubly-bottom tetraquark $T_{bb} = bb \bar u \bar d, \ I(J^P) = 0(1^+),$ has been particularly scrutinized as one the most promising candidates for an exotic QCD stable state \cite{Eichten:2017ffp,ref:bw_tbb,ref:tbb_quarkmodel1,PhysRevD.79.074010,Brambilla:2024imu,Hoffer:2024fgm,ref:rosina_tcc,Maiani:2022qze}. State-of-the-art lattice studies find its mass to be significantly below the lowest compatible decay threshold, $BB^*$, in the $I=0$ channel \cite{ref:hud_tbb,ref:bw_tbb,ref:tbb_leskovec,PhysRevD.110.094503,PhysRevD.110.054510,PhysRevD.108.054502,PhysRevD.99.034507,rlgp-c9tb,Tripathy:2025vao}. These studies are especially pertinent given the recent discovery of $T_{cc}=cc\bar{u}\bar{d}$ at LHCb~\cite{LHCb:2021vvq} and the favourable prospects of observing 
$T_{bc}=bc\bar{u}\bar{d}$ \cite{Ali:2018xfq,polyakov}, expected to lie slightly below the $DB^*$ threshold \cite{Alexandrou:2023cqg,Radhakrishnan:2024ihu,Padmanath:2023rdu}. 
Regarding $T_{bb}$, its production cross section at the LHC was determined to be a few $\mathrm{nb}$ \cite{Ali:2018xfq}. Detection in exclusive decays seems unlikely \cite{Gershon:2018gda}, while there may be a possibility of observing $T_{bb}$ in inclusive decays \cite{Gershon:2018gda}.\\
\indent Information about the structure of composite particles, i.e. their spin, color  and orbital wave functions, are encoded in their form factors. They appear via Lorentz-covariant parametrizations of matrix elements $\mathcal M = \langle h_2(p_2, \lambda_2) | \hat \jmath (x=0) | h_1(p_1, \lambda_1) \rangle$, with $h_{1(2)}$ being single hadrons possessing momenta and helicities $p_{1(2)}$ and $ \lambda_{1(2)}$, and  the current $\hat \jmath$ representing the probe. Currently, almost all theoretically or experimentally studied form factors of QCD states involve the conventional states, e.g. pions \cite{JeffersonLabFpi:2007vir,JeffersonLab:2008jve,PhysRevD.104.114515,PhysRevD.105.054502,CMD-3:2023rfe,PhysRevD.110.094505}, nucleons \cite{ref:P_and_N_EMFF,ref:N_EMFF,PhysRevD.105.054505,PhysRevD.105.094022,JeffersonLabHallA:1999epl,JeffersonLabHallA:2023rsh}, kaons \cite{Carmignotto:2018uqj,PhysRevD.105.054502,PhysRevD.110.094505} and others (including transition and multi-particle form factors) \cite{ref:NtoDelta1,ref:protonGFM1,Delaney:2023fsc,PhysRevD.73.074507,Abbott:2025irb,PhysRevD.103.074018,PhysRevC.55.448,PhysRevLett.134.161901}.\\ 
\indent Tetraquarks admit two color configurations that yield a $SU(3)$ singlet hadron \cite{PhysRevD.109.076026, Bicudo:2022cqi,Brambilla:2019esw}. One is the so-called {\it molecular structure} where the wave function  is  a product of two quark-antiquark pairs, both in definite color representations:    $(\bar q_1 q_3)_1  (\bar q_2 q_4)_1$, $(\bar q_1 q_4)_1  (\bar q_2 q_3)_1$ and $(\bar q_1 q_3)_8  (\bar q_2 q_4)_8$. The first two feature pairs of color-singlet hadrons 
interacting via residual color interactions. This structure prevails in nuclei composed of protons and neutrons,  and plays a prominent role for many tetraquark residing near meson-meson thresholds.  The second possibility includes the {\it diquark-antidiquark structure}: $[\bar q_1 \bar q_2]_{3} [q_3 q_4]_{\bar 3}$ or $[\bar q_1 \bar q_2]_{\bar 6} [q_3 q_4]_{6}$. The former features a diquark in a color-antitriplet  and an antidiquark in a color-triplet,  while the latter is comprised of a color-sextet and a color-antisextet. These color wave-functions are not linearly independent and can be related to each other \cite{PhysRevD.92.034501,PhysRevD.109.076026}. \\
\indent In this work, we probe the structure of the doubly-bottom tetraquark $T_{bb}$ by determining its electromagnetic form factors in lattice QCD. Its stability against strong decay guarantees that infinite-volume matrix elements $\langle T_{bb} | \hat \jmath | T_{bb} \rangle$ are directly accessible from the lattice and are not significantly polluted by finite-volume effects \cite{Briceno:2015tza,Baroni:2018iau}. This feature sets the $T_{bb}$ apart from all so-far discovered tetraquarks, which decay via strong interaction and therefore do not correspond to QCD asymptotic states. The $T_{bb}$, by contrast only decays weakly—like the pion, kaon, and other pseudoscalar hadrons. As such, it provides a unique opportunity to study the structure of a hadron, whose manifestly exotic character arises from its flavor content (two bottom quarks) and its electric charge.\\
\indent \textbf{\textit{$h \xrightarrow{\hat \jmath_{EM}^\mu} h$ form factor decomposition}}: The matrix element linked to the electromagnetic (EM) composition of a hadron $h$ is defined in the continuum as 
\begin{gather}
    \mathcal{M}_{EM}^\mu (p_2, \lambda_2, p_1, \lambda_1) =\langle h (p_2, \lambda_2) |\hat \jmath_{EM,cont}^\mu | h (p_1, \lambda_1) \rangle, \nonumber \\[1pt]
    \hat \jmath_{EM, cont}^\mu  = \sum_{q} e_q \bar q  \gamma^\mu q , \label{eq:current}
\end{gather}
where $e_q$ is the charge of quark $q$.\\ 
\indent The decomposition of matrix element \eqref{eq:current} is uniquely determined by the particle spin and the current $\hat \jmath$ \cite{PhysRev.126.1882}. For the same pseudoscalar in the initial and final state this renders a single form factor \cite{Khodjamirian:2020btr}
\begin{align}\label{eq:decomp_ps}
    \mathcal{M}_{EM}^\mu = (p_1 + p_2)^\mu F_C(Q^2),
\end{align}
while for $J^P = 1^\pm$ particles this gives \cite{PhysRevC.23.363}
\begin{align}\label{eq:decomp_vax}
    \mathcal{M}_{EM}^\mu =&-(p_1 + p_2)^\mu (\varepsilon_2^*  \cdot \varepsilon_1  ) F_1 (Q^2 ) - \\[1pt]
    &- [(\varepsilon_2^*  \cdot q)\varepsilon_1^\mu - (\varepsilon_1  \cdot q)\varepsilon_2^{*\mu}] F_2(Q^2) + \nonumber \\[1pt]
    &+ \frac{(\varepsilon_2^* \cdot q) (\varepsilon_1 \cdot q)}{2m^2} (p_1 + p_2)^\mu F_3(Q^2) \nonumber,
\end{align}
with $Q^2 \equiv -q^2 = -(p_2 - p_1)^2 > 0$ defined as the momentum transfer and $\varepsilon_{1(2)}^{(*)}$ as the polarization four-vectors. $F_{C,1,2,3} (Q^2)$ are Lorentz-scalars that are functions of $Q^2$. $F_{1,2,3} (Q^2)$ are further related to the charge, magnetic dipole and electric quadrupole form factors, labeled by $F_C, F_M$ and $F_{\cal{Q}}$, respectively, via a linear transformation \cite{PhysRevC.23.363,PhysRev.136.B140}
\begin{align}\label{eq:multipoleFF}
    \begin{pmatrix}
    F_C (Q^2) \\[1pt]
    F_M (Q^2) \\[1pt]
    F_{\cal{Q}} (Q^2)
    \end{pmatrix}
    =
    \begin{pmatrix}
    1+\frac{2}{3}\eta & - \frac{2}{3}\eta & \frac{2}{3}\eta(1+\eta) \\[1pt]
    0 & 1 & 0 \\[1pt]
    1 & -1 & (1 + \eta)
    \end{pmatrix}
    \begin{pmatrix}
    F_1 (Q^2) \\[1pt]
    F_2 (Q^2) \\[1pt]
    F_3 (Q^2)
    \end{pmatrix} ,
\end{align}
with $\eta = \tfrac{Q^2}{4m^2}$. The normalization of the charge form factor at $Q^2\! =\! 0$ is fixed to the total charge $Z$ of the state in units of the elementary charge $e_0$, i.e. $F_C (Q^2\! =\! 0)\! =\! Z$. The values of the electric quadrupole and the magnetic dipole form factors at $Q^2\!=\!0$ yield the values of the total electric quadrupole moment $\cal Q$ $=\frac{1}{m^2}F_{\cal Q} (0)$ and the magnetic dipole moment $\mu = \frac{1}{2m} F_M(0)$ \cite{PhysRevD.100.036008,Lorce:2009br}.\\
\indent \textbf{\textit{Lattice setup}}: We employ a single ensemble of gauge  configurations (\texttt{X253}) with a hypervolume $N_L^3 \!\times\! N_T \!=\! 40^3 \times 128$ and lattice spacing $a\!=\!0.06379(37) \ \mathrm{fm}$ \cite{ref:lattspacing1}, generated by the Coordinated Lattice Simulations (CLS) effort \cite{Bruno:2014jqa}. This $N_f \!=\! 2\!+\!1$ ensemble features $s$ and degenerate $u/d$ quarks described by the non-perturbatively $O(a)$ improved Wilson action, resulting in a pion mass $m_\pi \approx 290 \ \mathrm{MeV}$, listed in Table \ref{tab:r2ch}. The bottom quark action is implemented with an anisotropic Clover action \cite{Chen:2000ej}, tuned to reproduce the physical masses and continuum energy-momentum dispersions of $B$ and $B^*$ mesons on the ensemble \cite{ParticleDataGroup:2024cfk}.\\
\indent The light and heavy EM currents 
\begin{gather}
\hat \jmath_{u/d}^\mu= Z_{u/d}^V \times \left( \tfrac{2}{3}\bar u\gamma^\mu u-\tfrac{1}{3}\bar d \gamma^\mu d\right),\ \hat\jmath_{b}^\mu=Z_b^V \times \left( -\tfrac{1}{3}\bar b  \gamma^\mu b \right), \nonumber \\[1pt] \hat\jmath_{EM}^\mu=\hat\jmath_{  u/d}^\mu+\hat\jmath_{ b}^\mu, \label{eq:jEM}
\end{gather}
are non-perturbatively renormalized with factors $Z_{u/d}^V$ and $Z_b^V$ determined from $T_{bb}$ matrix elements to recover the infinite-volume and continuum normalization of the $T_{bb}$ charge form factor, $F_C(0)=-1$. The $b-$quark action, its tuning and current renormalization are described in the Supplemental Material \cite{ref:supplemental}.\\
\begin{figure}[t]
    \begin{overpic}[width=0.49\columnwidth]{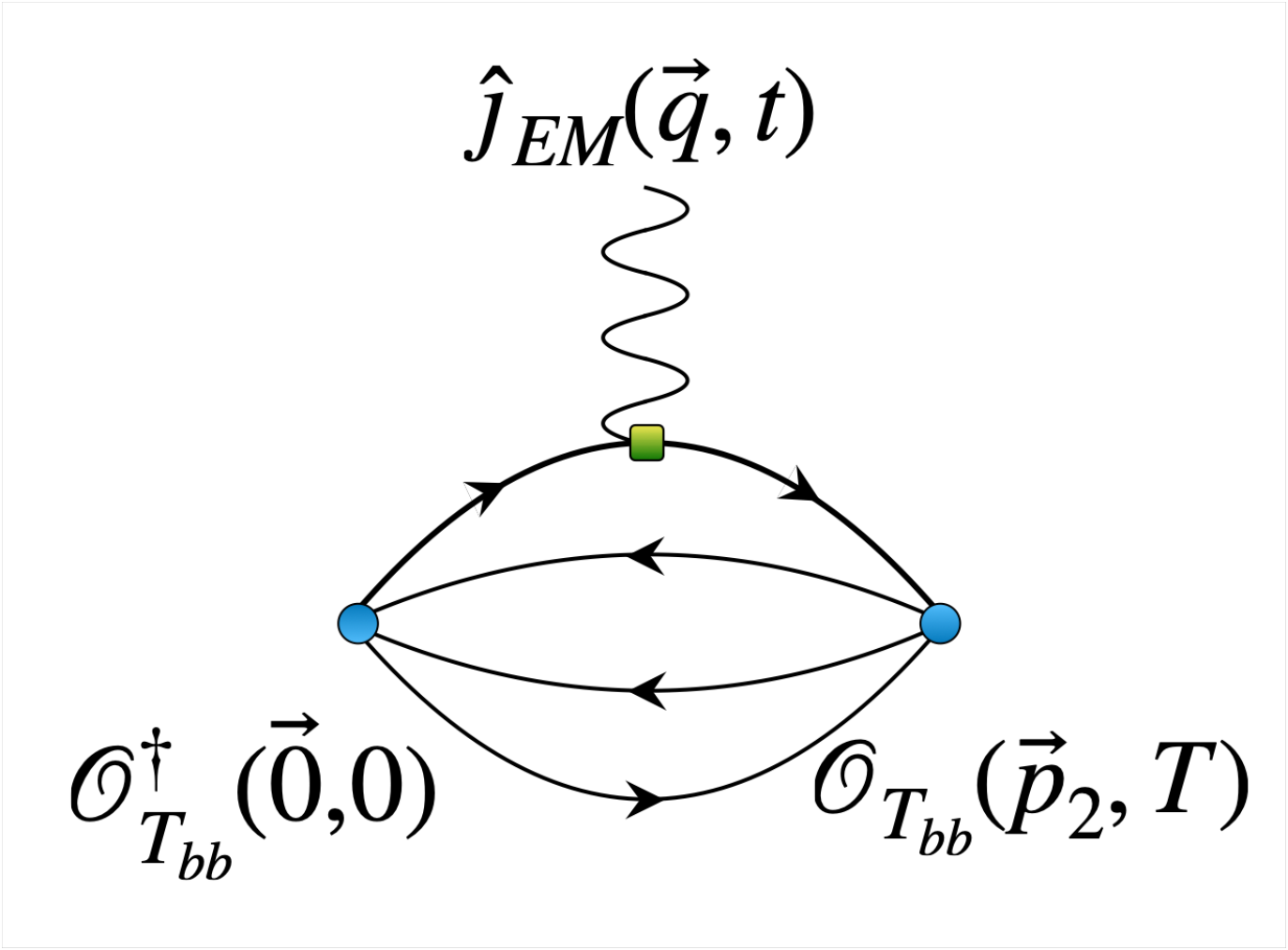}
    \put(0,71){\textbf{a)}}
    \end{overpic}
    \hfill
    \begin{overpic}[width=0.40\columnwidth]{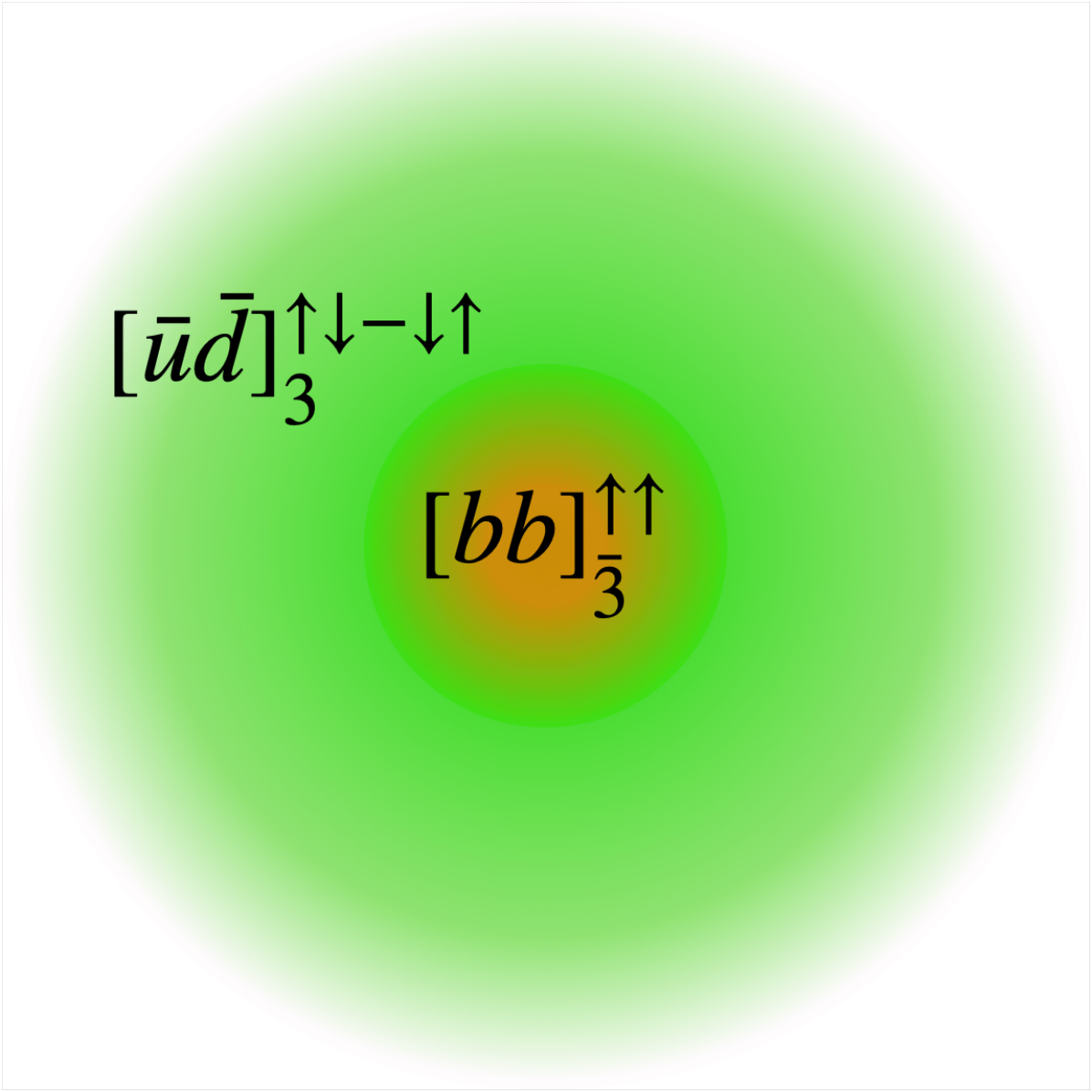}
    \put(0,90){\textbf{b)}}
    \end{overpic}
    \caption{\textbf{a)} Example of a connected Wick contraction diagram generated by the $T_{bb}$ three-point correlator \eqref{eq:threept_corr}. \textbf{b)} Pictorial representation of the resulting distributions for light and heavy quarks.  }
    \label{fig:threept}
\end{figure}
\indent  The desired matrix elements \eqref{eq:current} for states $h=T_{bb}, B, B^*, \pi$ (generated by the interpolators $O_h^{(\dagger)}$ as detailed in \cite{ref:supplemental}) were computed from three-point correlation functions $\mathcal{C}_3^\mu (\vec p_2, \vec q, T; t)$, shown for $T_{bb}$ in Fig.~\ref{fig:threept}:
\begin{align}\label{eq:threept_corr}
    \mathcal{C}_{3}^\mu &(\vec p_2, \vec q, T; t) = \langle \Omega | \mathcal{O}_{h} ( \vec p_2, T) \hat \jmath_{EM}^\mu (\vec q, t) \mathcal{O}^\dagger_{h} (  x=0) | \Omega \rangle = \nonumber \\[1pt]
    &= \sum_{n,m = 0}^\infty \frac{\mathcal{Z}^{f*}_n \mathcal{Z}^i_m}{(2E^f_n)(2E^i_m)}  \mathcal{M}_{nm}^\mu \ e^{-E^f_n (T-t)} e^{-E^i_mt}. 
\end{align}
Here $\vec p_2$ and $\vec q \equiv \vec p_2 - \vec p_1$ are momenta at the sink and the current, respectively, while  $T$ denotes the source-sink temporal separation. The three-point function is decomposed in terms of initial states $(i,m)$ and final states $(f,n)$ in Euclidean time. The $\mathcal{Z}_{a} = \langle a | \mathcal{O}^\dagger_h | \Omega \rangle$  labels the overlap with the $a$-th state and $E_a$ denotes its energy. In addition, operators $\mathcal{O}_h$ for $h=B^*,T_{bb}$ with nonzero spin are projected to the appropriate rows $r$ and irreps $\Lambda$ of subgroups of the octahedral group. In addition to the connected Wick contraction in Fig.~\ref{fig:threept}\textbf{a}, there might be in principle also a nonvanishing contribution of disconnected diagrams. The latter are typically found to be very small for EM currents (e.g. \cite{Green:2015wqa} for nucleon) and we omit them in the current study.\\
\begin{figure*}[t]
    \begin{overpic}[width=0.47\textwidth]{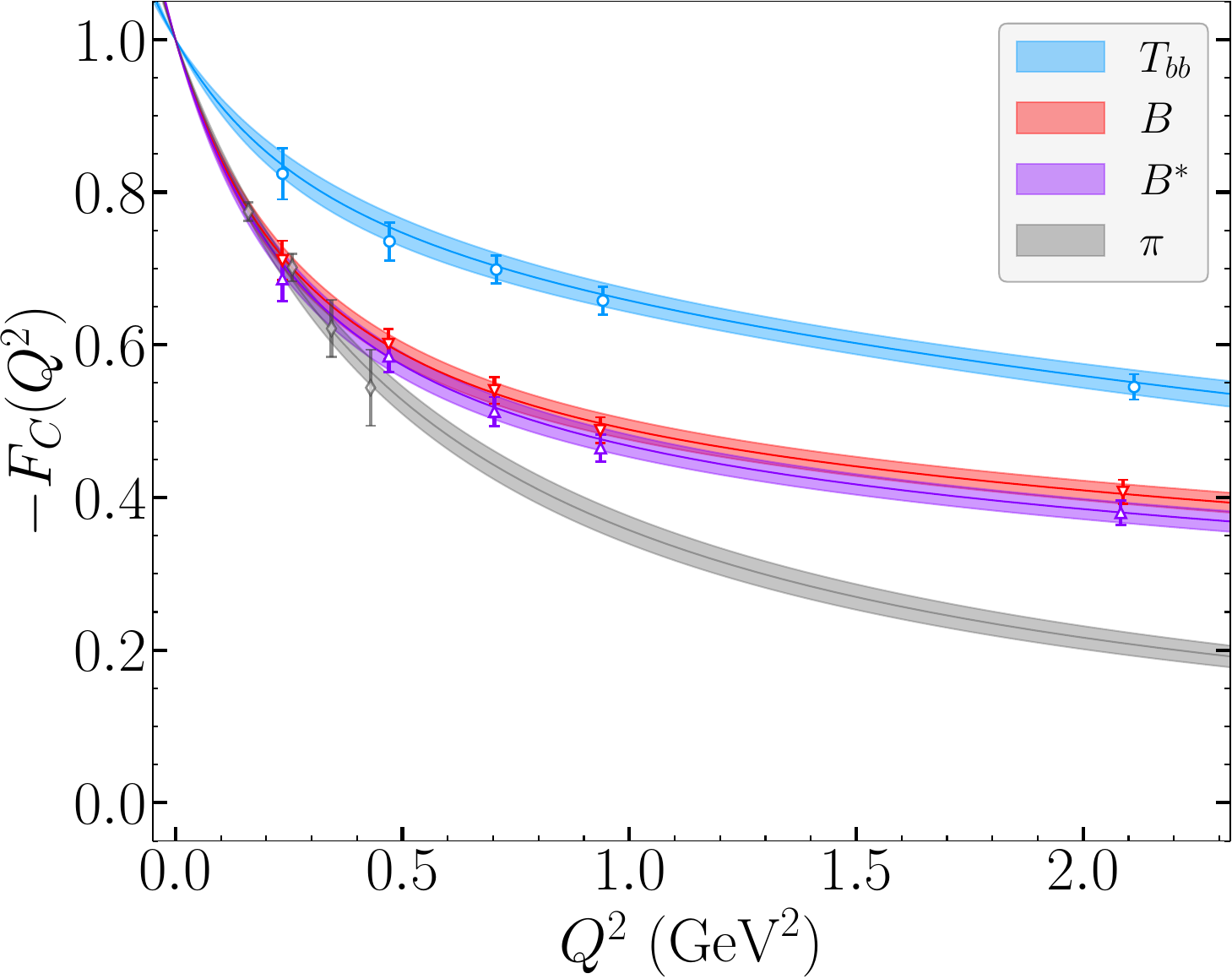}
        \put(0,79){\textbf{a)}}
    \end{overpic}
    \hfill
    \begin{overpic}[width=0.47\textwidth]{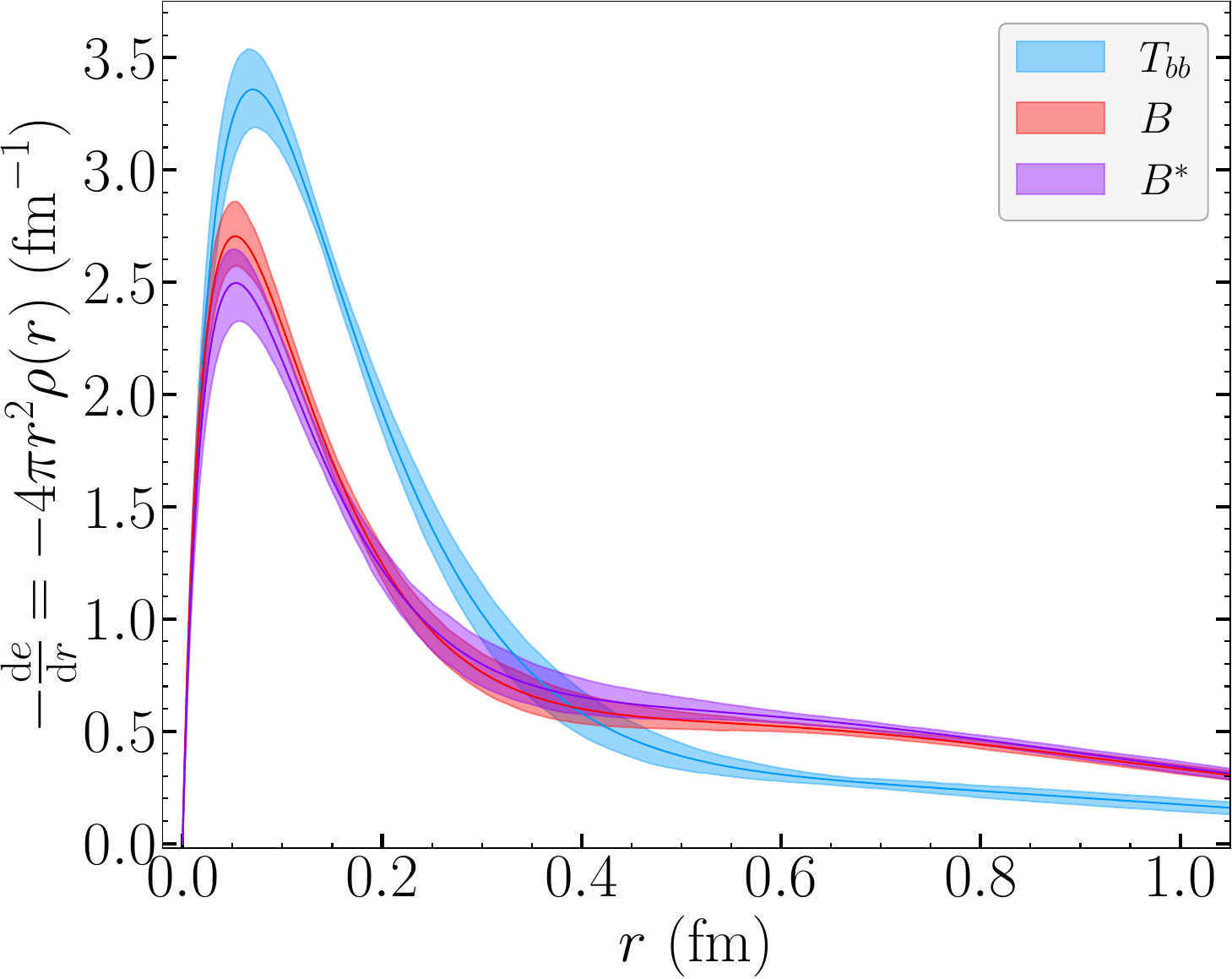}
        \put(0,79){\textbf{b)}}
    \end{overpic} 
  \caption{\textbf{a)} Charge form factors $F_C (Q^2)$ of hadrons $h = T_{bb}, B(b\bar u), B^*(b\bar u), \pi(d \bar u)$, shown as a function of $Q^2$. Discrete markers show lattice data, while the bands represent $z-$expansion fits. Second order expansion (up to and including $n\!=\!2$ in eq.~\eqref{eq:zexp}) was used to parametrize $T_{bb}, \ B, \ B^*$ electric form factors, while a first order expansion sufficed for an adequate parametrization of the pion form factor. \textbf{b)} Position-space charge densities $\rho(r)$, represented in the form of $-\tfrac{\mathrm{d}e}{\mathrm{d}r}=-4\pi r^2\rho$, are related to the form factors via a Fourier transform.  Negative values of charge form factors and distributions are shown, given that considered hadrons are negatively charged.}
  \label{fig:fc_comp}
\end{figure*}
To isolate the ground-state matrix element $\mathcal{M}^\mu_{00} \equiv \mathcal{M}^\mu$, we remove the zeroth-order dependence of the three-point correlator \eqref{eq:threept_corr} on the ground-state overlap factors $\mathcal{Z}_0^{f(i)}$ and energies $E^{f(i)}_0$, obtained from the corresponding two-point correlators $C_2 (\vec p, t) = \langle \Omega | \mathcal{O}_h (\vec p, t) \mathcal{O}^\dagger_h (0) | \Omega \rangle$. This is done by constructing the ratio $R_3^\mu (\vec p, \vec q, T; t)$
\begin{align}\label{eq:ratio3pt}
    R_3^\mu (\vec p_2, \vec q, T; t)  = \frac{(2E^f_0)(2E^i_0)}{\mathcal{Z}^{f*}_0 \mathcal{Z}^i_0}e^{E^f_0(T-t)} e^{E^i_0 t}C_3^\mu (\vec p_2, \vec q, T; t),
\end{align}
defined in eq. (28) of Supplemental Material of Ref. \cite{PhysRevLett.134.161901},
that equals the desired matrix element, up to excited-state contamination. Two models are used in fits to data: a constant fit, assuming no contamination, or a model that also incorporates first-excited states \cite{ref:supplemental}. To improve fit quality, we performed a weighted average of the ratios \eqref{eq:ratio3pt} over selected equivalent directions of the sink and source momenta that yield the same value of $Q^2$. The averaged data was simultaneously fitted for four source-sink separations $\tfrac{T}{a} = 12, 15, 18, 22$. The averaging procedure and plots showing representative fits are given in the Supplemental Material \cite{ref:supplemental}. \\
\begin{table}[h!]
  \centering
  \begin{tabular}{l l l r c}
    \toprule
    $h$ & $m_h$ (GeV) & $\sqrt{\langle r_C^2 \rangle}$ ($\mathrm{fm}$) & $r$ & $t_+$ \\
    \midrule
    $T_{bb}$ & $10.5765(98)$ & $0.499(31)$ & $\omega$ & $(3m_\pi)^2$   \\
     $B$ & $5.3020(17)$& $0.692(21)$ & $\rho$ & $(2m_\pi)^2$ \\
     $B^*$ & $5.3387(20)$& $0.698(23)$ & $\rho$ & $(2m_\pi)^2$ \\
     $\pi$ & $0.28953(97)$& $0.652(20)$ & $\rho$ & $(2m_\pi)^2$ \\
    \bottomrule
  \end{tabular}
  \caption{Hadron masses $m_h$ and charge radii $\sqrt{\langle r_C^2 \rangle}$. Columns $r$ and $t_+$ show the resonances and multi-particle thresholds, respectively, that appear in the $z-$expansions.}
  \label{tab:r2ch}
\end{table}

\begin{figure*}[t]
    \begin{overpic}[width=0.325\textwidth]{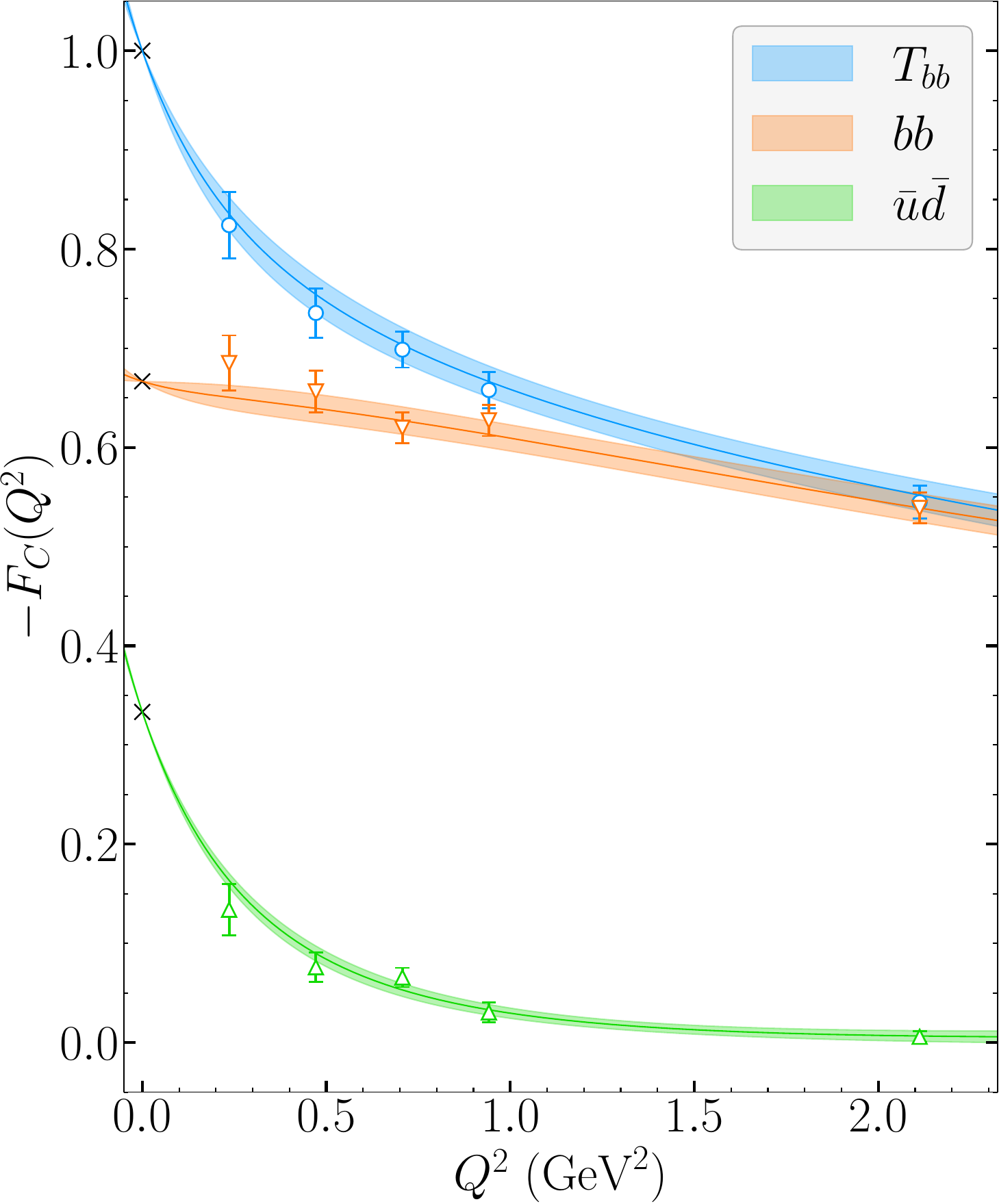}
    \put(0,99){\textbf{a)}}
    \end{overpic}
    \hfill
    \begin{overpic}[width=0.325\textwidth]{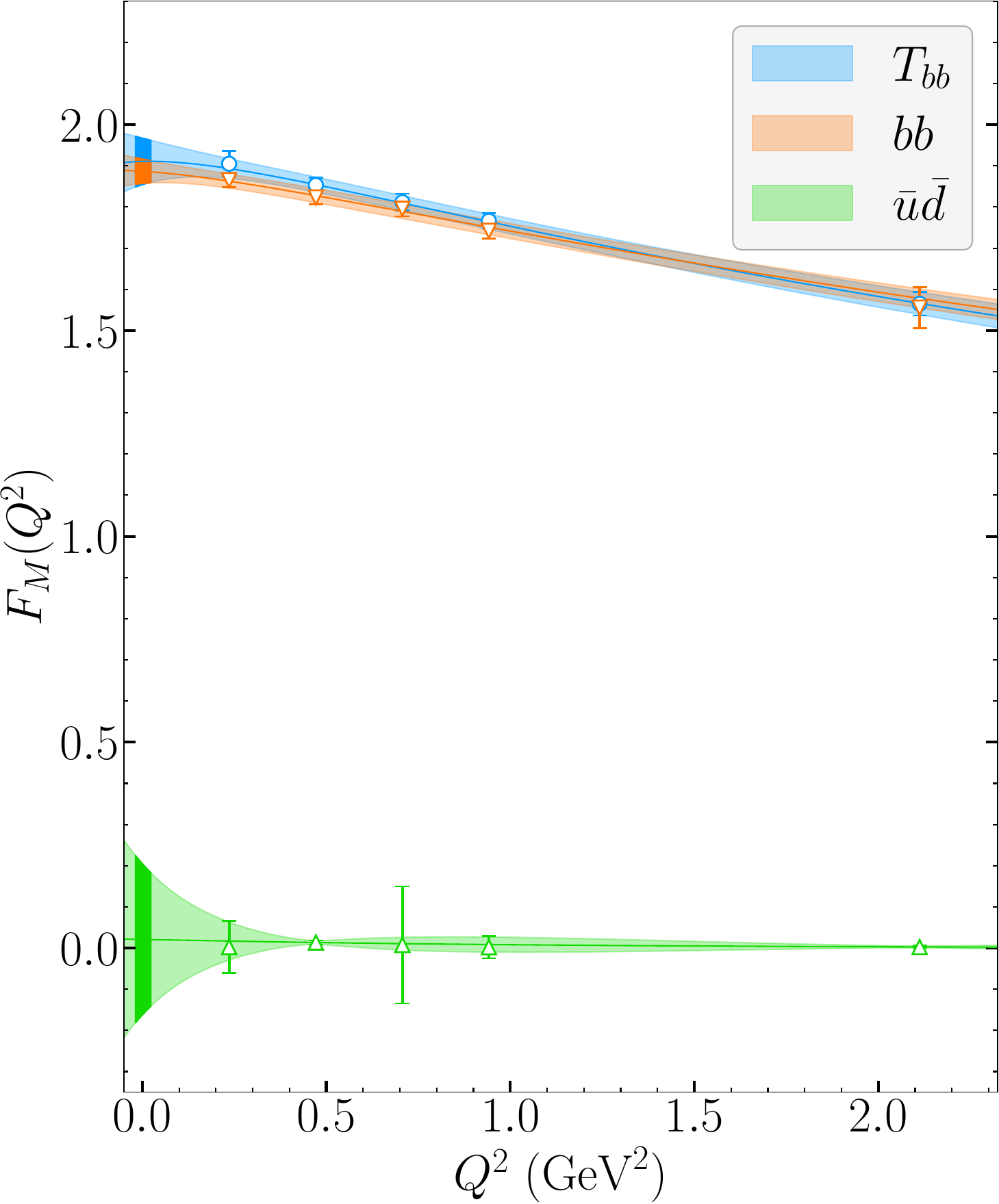}
    \put(0,99){\textbf{b)}}
    \end{overpic} 
    \hfill
    \begin{overpic}[width=0.325\textwidth]{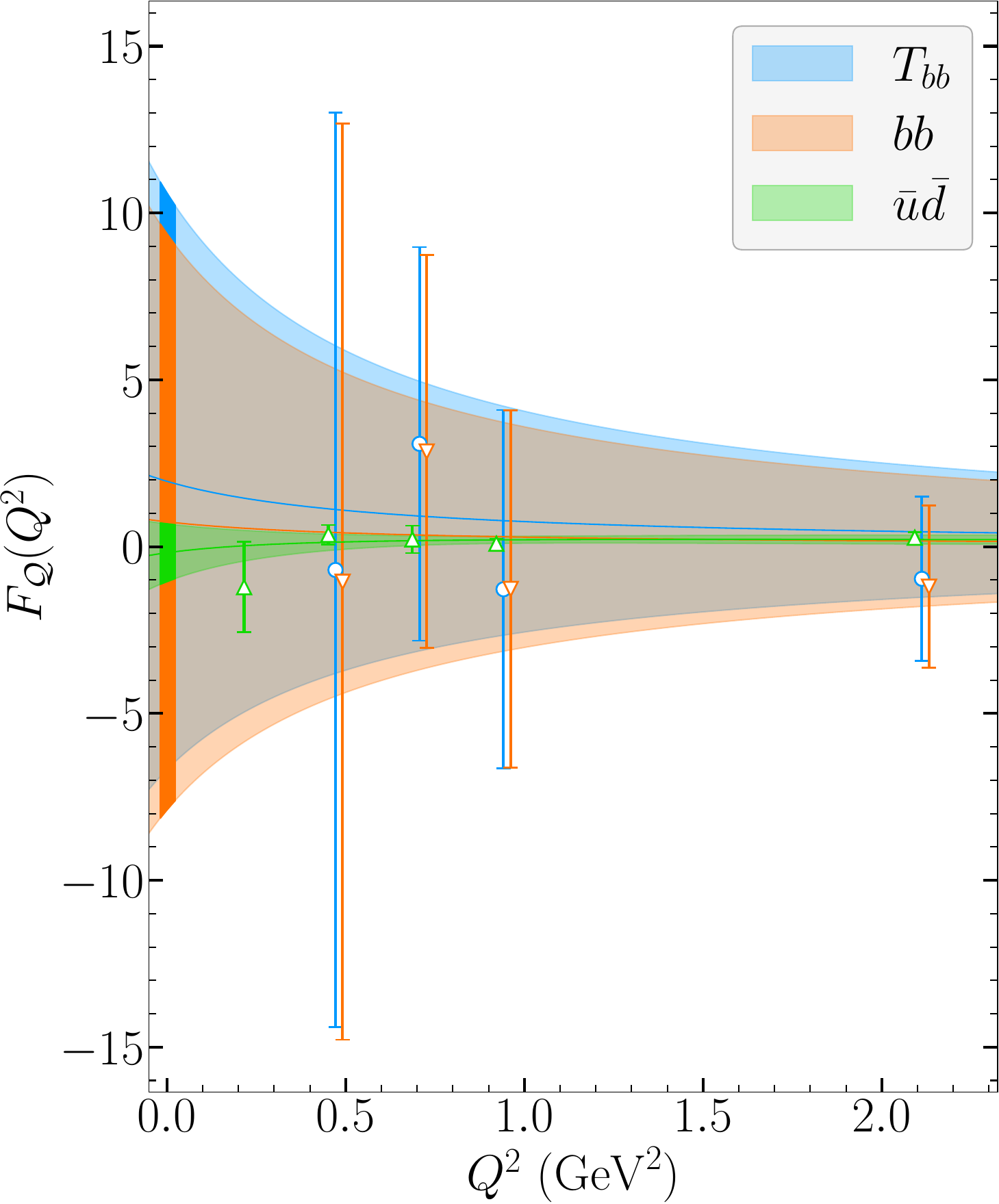}
    \put(0,99){\textbf{c)}}
    \end{overpic} 
    \caption{Form factors of the $T_{bb}$. Each subfigure shows the total value of the form factors with separate contributions yielded by the light current 
  $\hat \jmath_{u/d}^\mu$ and the heavy current $\hat \jmath_b^{\mu}$. The crosses in \textbf{a)} mark values to which each of the charge form factors have been normalized at $Q^2=0$, while the shaded vertical bands in \textbf{b)} and \textbf{c)} indicate the values of magnetic dipole moments $2m_{T_{bb}}\cdot \mu$ and electric quadrupole moments $m_{T_{bb}}^2 \cdot \cal Q$, respectively, also found in Table \ref{tab:tbb_moments}. The points on the rightmost plot are slightly horizontally displaced for improved visibility.}
    \label{fig:tbb_formfac}
\end{figure*}

\indent \textbf{\textit{Results}}: The extracted masses of $T_{bb}$, $B$ and $B^*$, listed in Table \ref{tab:r2ch}, render a significant binding energy of the $T_{bb}$ with respect to the $BB^*$ threshold
\begin{equation}
m_{T_{bb}} - (m_{B} + m_{B^*}) = -64(10) \ \mathrm{MeV},
\end{equation}
at $m_\pi \approx 290 \ \mathrm{MeV}$. This is in line with the large binding observed in previous studies, e.g.  \cite{ref:hud_tbb,ref:bw_tbb,ref:tbb_leskovec,PhysRevD.110.094503,PhysRevD.110.054510,PhysRevD.108.054502,PhysRevD.99.034507,Tripathy:2025vao,Brambilla:2024imu}.\\
\indent The EM form factors  
of hadrons $h=T_{bb},\ B,\ B^*,\ \pi$ are shown in Figs.~\ref{fig:fc_comp}\textbf{a} and \ref{fig:tbb_formfac} for five values of $Q^2$. The latter plot also shows the individual EM form factors of the light and heavy currents, extracted from matrix elements $\langle T_{bb} | \hat \jmath_{u/d}^\mu | T_{bb} \rangle$ and $\langle T_{bb} | \hat \jmath_{b}^\mu | T_{bb} \rangle$, respectively.\\
\indent Continuous bands  in Figs.~\ref{fig:fc_comp} and \ref{fig:tbb_formfac} follow from parametrizing all form factors using the $z-$expansion \cite{Boyd:1994tt,Boyd:1997qw,Bourrely:2008za} of the form:
\begin{align}\label{eq:zexp}
    F(Q^2) = \frac{1}{1 + \frac{Q^2}{m_r^2}} \sum_{n} a_n z^n (Q^2; t_+, t_0).
\end{align}
Here $m_r$ is the mass of the closest resonance and $z$ is a variable that maps $Q^2$ to the unit disk, 
\begin{align}\label{eq:zvar}
    z(Q^2; t_+, t_0) = \frac{\sqrt{t_+ + Q^2} - \sqrt{t_+ - t_0}}{\sqrt{t_+ + Q^2} + \sqrt{t_+ - t_0}},
\end{align}
where $t_+$ is the squared mass of the nearest multi-particle threshold and $t_0$ is a tunable parameter. Table \ref{tab:r2ch} lists the closest resonances and thresholds employed in fitting eq.~\eqref{eq:zexp} to all form factors. The thresholds $t_+$ are evaluated at $m_\pi \approx 290 \ \mathrm{MeV}$. As the $\rho$ and $\omega$ resonance masses are not known on our ensemble, we use the PDG values and verify that the fits are robust when varying these masses within the range $m_{\omega,\rho} \in [ 750 \ \mathrm{MeV}, 900 \ \mathrm{MeV} ]$. The coefficients $a_n$ are fit parameters, and we truncate all expansions at order $n=2$ or less, with their numerical values, and that of $t_0$, given in the Supplemental Material \cite{ref:supplemental}.\\   
\indent The electric charge and quadrupole form factors encode the information about the spatial charge distribution $\rho (\vec r)$ \cite{Lorce:2009br} and consequently the charge radius $r_C \equiv \sqrt{\langle r_C^2 \rangle}=\sqrt{6\cdot\tfrac{\mathrm{d}F_C}{\mathrm{d}Q^2} (0)}$. We find that the quadrupole contribution in $T_{bb}$ is negligible with respect to the monopole contribution, as shown in Fig. 3 of \cite{ref:supplemental}. The corresponding monopole of the spatial charge distribution is obtained via the Fourier transform, $\rho(r)= \int \tfrac{\mathrm{d}^3  q}{(2\pi)^3} ~e^{i\vec q \cdot \vec r} F_C(|\vec q|^2)  $, in the non-relativistic limit, which a good approximation for $T_{bb},\ B$ and $B^*$. The resulting charge radius of the $T_{bb}$, $0.499(31)$ fm, is smaller than that of the $B$ or $B^*$ individually (listed in Table \ref{tab:r2ch}), and significantly smaller than the sum of both, $1.390(31)$ fm. This is consistent with the $T_{bb}$ charge density being concentrated at a smaller radius $r$ than the charge densities of $B$ or $B^*$, see Fig.~\ref{fig:fc_comp}\textbf{b}. The compactness of the $T_{bb}$ implies that the notion of a $B-B^*$ molecule is not meaningful and instead favors a compact diquark-antidiquark composition \cite{rlgp-c9tb,ljt6-cv33,Cheng:2020wxa,Lee:2009rt}. $B$ meson charge radius we obtain is comparable to the values found in Refs. \cite{Hwang:2001th,Miramontes:2025vzb} and has been studied previously on the lattice through radial charge distributions in heavy-light mesons \cite{Becirevic:2009ya,PhysRevD.65.014512,Koponen:2002sc} with static heavy quarks.  \\
\indent The diquark-antidiquark structure is further supported by investigating the charge distributions of light and heavy (anti)diquarks separately in Fig.~\ref{fig:tbb_formfac}\textbf{a}. We find that the $[bb]$ diquark component of the total charge form factor has significantly weaker $Q^2$ dependence than the antidiquark $[\bar u \bar d]$ component, indicating that the diquark is more localized than the antidiquark, see Table \ref{tab:tbb_moments}. This suggests that the structure with a very compact heavy diquark and a relatively spatially extended antidiquark is responsible for the large binding energy of the tetraquark, compatible with some quark model studies \cite{PhysRevC.75.045206,Luo:2017eub,Cheng:2020wxa,PhysRevD.79.074010,PhysRevD.109.076026,ljt6-cv33,PhysRevD.101.014001}.\\
\begin{table}[h!]
  \centering
  \begin{tabular}{c c c c}
    \toprule
        {} & $T_{bb}$  & $[ bb ]$ & $[ \bar u \bar d ]$ \\
    \midrule
    $\sqrt{\langle r_C^2 \rangle} \ (\mathrm{fm})$ & $0.499(31)$ & $0.174(59)$ & $0.511(14)$  \\
    $2m_{T_{bb}} \cdot \mu$ & $1.912(57)$ & $1.887(29)$ & $0.02(18)$   \\
     $m_{T_{bb}}^2 \cdot \cal Q$ & $1.9(8.7)$& $0.6(8.8)$ & $-0.18(86)$ \\ 
    \bottomrule
  \end{tabular}
  \caption{Charge radii, magnetic dipole $\mu$ and electric quadrupole moments ${\cal Q}$ of $T_{bb}$ and the constituent (anti)diquarks. The same values are also shown with shaded bands at $Q^2=0$ in Figs.~\ref{fig:tbb_formfac}\textbf{b} and \ref{fig:tbb_formfac}\textbf{c}.}
  \label{tab:tbb_moments}
\end{table}
\indent Motivated by our findings, we assume that the $T_{bb}$ state vector in the QCD Hilbert space can be expanded in the diquark-antidiquark basis with compatible quantum numbers, in line with quark models applied to QCD exotics. Each diquark-antidiquark state is factorized into three components~(orbital, spin and color)~\cite{PhysRevD.109.076026,PhysRevD.79.074010}:
\begin{align}\label{eq:QQqq}
    { \{[bb]_{\bar c}^{l_{bb},s_{bb}}[\bar u\bar d]_{c}^{I=0,l_{\bar u \bar d},s_{\bar  u \bar d}} \}^{J^P=1^+}_{l_{12}} } 
\end{align}
where $l_{bb}$ and $l_{\bar u \bar d}$ denote the relative orbital angular momenta in the diquark and antidiquark, respectively, and $s_{bb}, s_{\bar u \bar d} = 0,1$ are their corresponding spins. $l_{12}$ is the relative angular momentum between the diquark and the antidiquark, whereas $c = 3, \bar 6$ are color configurations, as discussed in the introduction. The values introduced in eq.~\eqref{eq:QQqq} are subject to the constraints imposed by the $T_{bb}$ quantum numbers $I(J^P) = 0(1^+)$ and the Pauli principle, ensuring that the total $T_{bb}$ state is antisymmetric with respect to the permutations of the $b$ quarks or the light antiquarks in the isospin limit. Accounting for this yields three relations \cite{PhysRevC.75.045206,PhysRevD.79.074010}
\begin{gather}
      (a) \hspace{0.2cm}  (-1)^{s_{bb} + l_{bb} + c} = 1, \hspace{0.5cm} (b) \hspace{0.2cm}  (-1)^{s_{\bar u \bar d} + l_{\bar u \bar d} + c} = -1 ,\nonumber \\[1pt]
 (-1)^{l_{bb} + l_{\bar u \bar d} + l_{12}} = 1 \ \overset{(a),(b)}{\longrightarrow} \  (-1)^{s_{bb} + s_{\bar u \bar d} + l_{12}} = -1 ,\label{eq:pauli}
\end{gather}
where the first row follows from the Pauli principle and the second ensures positive parity.\\
\indent Our results for the electric quadrupole form factors are consistent with zero, as presented in Fig.~\ref{fig:tbb_formfac}\textbf{c} and Table \ref{tab:tbb_moments}, given the present level of accuracy. This applies to the total value and separate diquark and antidiquark contributions. It indicates that the $S-$wave $(l_{bb} = l_{\bar u \bar d} = l_{12} = 0)$ components dominate in the relative wave functions within the diquark, antidiquark, and the total system \cite{Amado:1981ura,Bashkanov:2019mbz}. The magnetic dipole moment of the $T_{bb}$
\begin{equation}
\mu_{T_{bb}} = \langle T_{bb} | \sum_{q=u,d,b} \frac{e_q}{2m_q} (\hat l_q + g_q \hat s_q) | T_{bb}\rangle=\frac{F_M(0)}{2m_{T_{bb}}},
\end{equation}
is determined from the value of the magnetic dipole form factor $F_M$ at $Q^2\!=\!0$. It is nonzero and almost completely saturated by the contribution from the heavy quarks as seen in Fig.~\ref{fig:tbb_formfac}\textbf{b}, implying that the light quarks dominantly form a spin singlet state, while the heavy quarks are in a spin triplet state
\begin{equation}
\label{spin}
s_{\bar u \bar d} = 0, \quad s_{bb} = 1.
\end{equation}
Note that $T_{bb}$ with molecular structure $B-B^*$  would not render a spin correlation within two light quarks, or correlation within two heavy quarks, disfavoring the molecular structure.
Taking into account the orbital and spin quantum numbers, eq.~\eqref{eq:pauli} restricts the diquark-antidiquark to be in an antisymmetric color triplet-antitriplet configuration, i.e. $c=3$. This uniquely determines the $T_{bb}$ state in QCD Hilbert space to be 
\begin{align}\label{eq:tbbvec}
| T_{bb} \rangle  = { \{[bb]_{\bar 3}^{l_{bb}=0,s_{bb}=1}[\bar u\bar d]_{3}^{I=0,l_{\bar u \bar d}=0,s_{\bar  u \bar d}=0} \}^{J^P=1^+}_{l_{12}=0} }, \nonumber \\[1pt]
\end{align}
which is the simplest and the dominant configuration not in tension with our lattice results. Result \eqref{eq:tbbvec} is graphically summarised in Fig.~\ref{fig:threept}\textbf{b}. \\
\indent The identified structure features  the “good” light antidiquark \cite{Jaffe:2004ph}, in which the attraction between two quarks increases with decreasing $m_{u/d}$ \cite{Francis:2021vrr}, therefore the dominance of this structure is expected to pertain also at the physical light quark masses. 
We also expect that our findings, including the assignment of the quantum numbers in (\ref{eq:tbbvec}), are not significantly affected by finite-volume or lattice spacing effects.
Future lattice QCD studies at smaller pion masses and lattice spacings, as well as  larger volumes are desirable to confirm this. \\
\indent \textbf{\textit{Conclusions}}: In this work we have presented the first lattice QCD calculation of electromagnetic form factors of an exotic tetraquark $T_{bb}$. All computations were done on one CLS ensemble at the pion mass $m_\pi\! \approx\! 290 \ \mathrm{MeV}$, at which we observe $T_{bb}$ to be a strongly stable state with a binding energy $m_{T_{bb}} - (m_{B} + m_{B^*}) = -64(10) \ \mathrm{MeV}$.\\
\indent By probing $T_{bb}$ with the electromagnetic light and heavy currents, we extracted three form factors that reveal three components of its structure: the electric charge distribution, the magnetic dipole moment, and the electric quadrupole moment. Flavor decompositions of all form factors were determined from the respective matrix elements.\\
\indent These observables were combined with constraints from the Pauli principle. We have shown that, within the precision of our data, the $T_{bb}$ is a compact bound state consisting of a heavy diquark $[bb]$ in a color-antitriplet with spin one, and a light antidiquark $[\bar u \bar d]$ in a color-triplet with spin zero. The structure of  $T_{bb}$  is therefore rather unique, and differs from a molecular structure that dominates in many tetraquarks  near meson-meson thresholds, and prevails in nuclei composed of protons and neutrons.\\
\indent In the future, it would be valuable to investigate the structure of the $\Lambda_b$ and $\Xi_{bb}$ baryons, which might be related to that of $T_{bb}$ and the $B$ meson, respectively, via the symmetry between the heavy diquark and the heavy antiquark. Similarly, determining the EM form factors of $T_{bc}$ and eventually $T_{cc}$, would shed light on the heavy quark mass dependence of
the internal configuration of double heavy tetraquarks.
\\
\indent \textit{Data availability}: Lattice data shown in this work can be found at \cite{dataavail}.
\\
\indent \textit{Acknowledgements}: We thank M. Padmanath and R. J. Hudspith for valuable discussions, especially during the early stages of the project regarding the tuning of the $b-$quark action. We thank our colleagues who are part of CLS for their joint effort in the generation of gauge configurations that were employed in lattice simulations for this study. Software packages QDP-JIT \cite{Winter:2014npa} and Chroma \cite{Edwards:2004sx} were used for computing the Wick contractions and inverting the heavy quark Dirac operator, while the QUDA multigrid solver was employed for the light quark Dirac operator inversions \cite{Clark:2009wm}. The authors gratefully acknowledge the HPC RIVR consortium (\href{www.hpc-rivr.si}{https://www.hpc-rivr.si}) and EuroHPC JU (\href{eurohpc-ju.europa.eu}{https://eurohpc-ju.europa.eu}) for funding this research by providing computing resources of the HPC system Vega at the Institute of Information Science (\href{www.izum.si}{https://www.izum.si/en/home}), in particular the project QCD on Vega (S24O01-37 and 525002-11). The authors also acknowledge the scientific support and HPC resources provided by the Erlangen National High Performance Computing Center (NHR@FAU) of the Friedrich-Alexander-Universität Erlangen-Nürnberg (FAU) under the NHR project b124da. NHR funding is provided by federal and Bavarian state authorities. NHR@FAU hardware is partially funded by the German Research Foundation (DFG) – 440719683.  The work of I. V., L. L. and S. P. is supported by the Slovenian Research Agency (research core Funding No. P1-0035 and J1-3034 and N1-0360).

\nocite{Chen:2000ej,ParticleDataGroup:2024cfk,Dudek:2010wm,Bernard:2008ax,PhysRevLett.134.161901,Boyd:1994tt,Boyd:1997qw,Bourrely:2008za,Lorce:2009br,APE:1987ehd,ref:stout,El-Khadra:1996wdx}

\bibliography{refs}

\newpage

\onecolumngrid

\newcounter{supsec}
\renewcommand{\thesupsec}{\Roman{supsec}}
\newcommand{\supsection}[1]{
  \refstepcounter{supsec}
  \section*{\thesupsec. #1}
  \label{sec:#1}
}

\renewcommand{\thetable}{S\arabic{table}}
\renewcommand*{\theHtable}{\thetable}  
\renewcommand{\thefigure}{S\arabic{figure}}
\renewcommand*{\theHfigure}{\thefigure}  
\renewcommand{\theequation}{S\arabic{equation}}
\renewcommand*{\theHequation}{\theequation}

\setcounter{table}{0}
\setcounter{figure}{0}
\setcounter{equation}{0}

\section*{Supplemental Material: Electromagnetic form factors and structure of the $T_{bb}$ tetraquark from lattice QCD}

\begin{figure*}[t]
  \centering
  \includegraphics[width=0.85\textwidth]{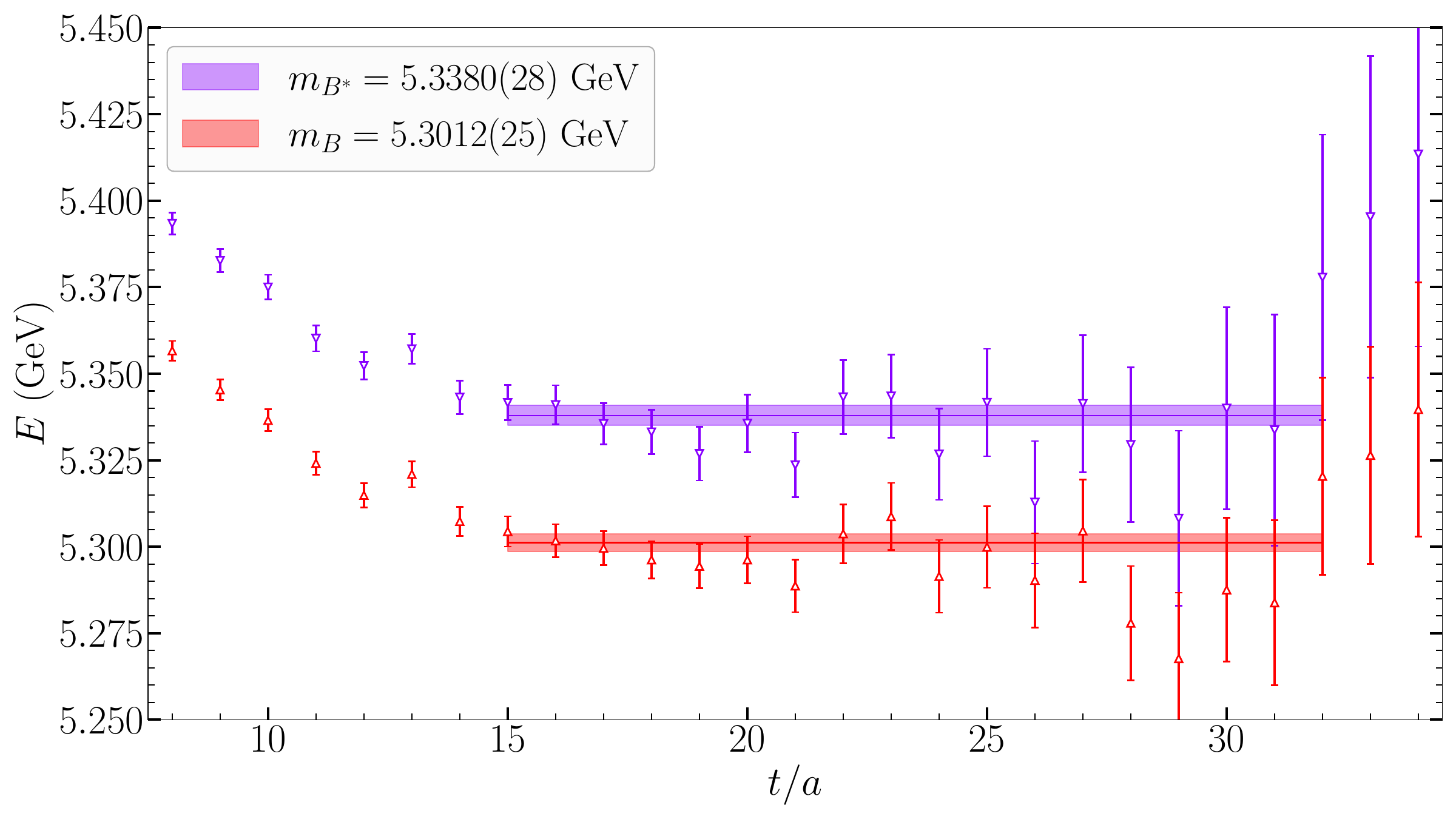}
  \caption{Plot showing the effective masses $m_{eff}(t) = -\frac{\ln (C_2(t+1))}{\ln(C_2(t))}$ of the $B$ and $B^*$. The mass, extracted from a single-exponential fit  in the plateau region $14a < t <33a$ of the $B^{(*)}$ two-point correlators, is shown with a line and error bands.} 
  \label{fig:meff_plot}
\end{figure*}

\begin{figure*}[t]
    \begin{overpic}[width=0.49\textwidth]{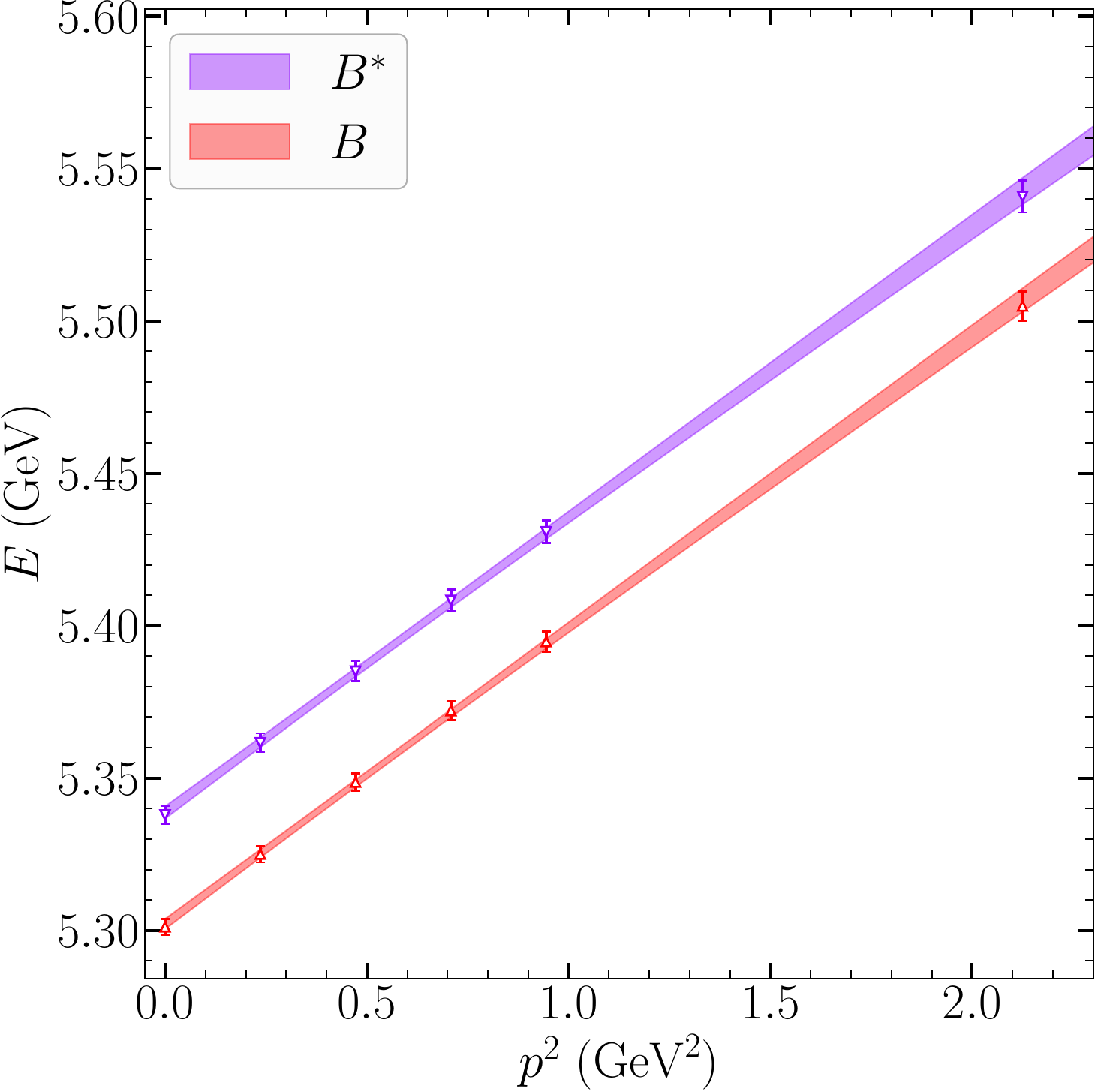}
    \put(8,105){\textbf{a)}}
    \end{overpic}
    \begin{overpic}[width=0.49\textwidth]{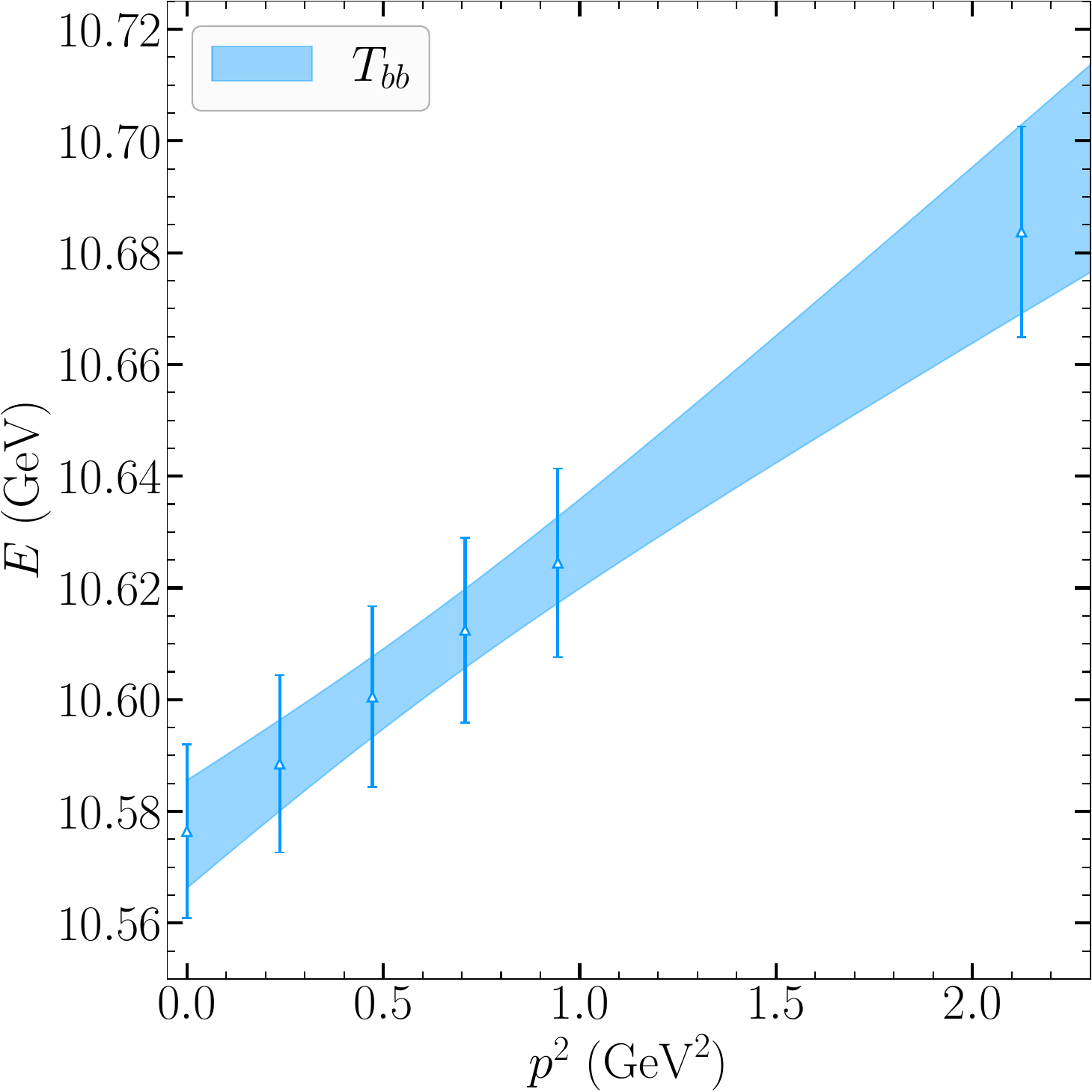}
    \put(8,105){\textbf{b)}}
    \end{overpic} 
    \caption{Final energy-momentum dispersion relations obtained using the tuned $b-$quark action \eqref{eq:action}. \textbf{a)} Dispersion relations of the $B$ and $B^*$ mesons. \textbf{b)} Dispersion relation of the $T_{bb}$.}
    \label{fig:disprels}
\end{figure*}

\supsection{Ensemble information}
The basic information about the employed ensemble  is listed in Table \ref{tab:ensemble_info}, where $a$ is the lattice spacing, $aN_L$ is the spatial extent, $aN_T$ is the time extent and $N_f$ is the number of dynamical quark flavors. $N_{conf} (B,B^*,T_{bb})$ shows the number of gauge configurations that was used in computations of the two- and three-point correlation functions for each of the hadrons. To estimate statistical uncertainties, we used jackknife resampling throughout this work.  
\begin{table}[h!]
  \centering
  \begin{tabular}{l c}
    \texttt{X253} & $\beta = 3.55$ \\
    \toprule
    $N_L$ & $40$  \\
    $N_T$ & $128$ \\
    $a$ & $0.06379(37)$ fm   \\
    $L$ & $2.552(15)$ fm  \\
    $N_f$ & 2+1 \\
    $m_\pi$ & $289.53(97)$ MeV \\
    $N_{conf} (B,B^*)$ & $996$ \\
    $N_{conf} (T_{bb})$ & $995$ \\ 
    \bottomrule
  \end{tabular}
  \caption{Summary of basic information about the CLS ensemble \texttt{X253} that was used in computations.}
  \label{tab:ensemble_info}
\end{table}
\supsection{Tuning of the bottom quark action}
To simulate the heavy $b-$quark on the lattice, we employed the Clover-improved Wilson action \cite{El-Khadra:1996wdx,Chen:2000ej}
\begin{align}\label{eq:action}
    \mathcal{S}_{WC}^Q = a^4 \sum_{x} \bar Q(x)  \left( m_Q + \gamma_t \nabla_t - \frac{1}{2}\Delta_t + \nu \sum_{i=1}^3 \left( \gamma_i \nabla_i - \frac{1}{2}\Delta_i \right) - \frac{c_E}{2} \sum_{i=1}^3 \sigma_{it}F_{it} - \frac{c_B}{2} \sum_{i<j} \sigma_{ij} F_{ij} \right) Q(x), 
\end{align}
with $\nabla_\mu, \ \Delta_\mu$ defined as the covariant first- and second-order discretized derivatives, respectively, as in \cite{Chen:2000ej} and $m_Q, \ \nu, \ c_E$ and $c_B$ serving as tunable parameters. Our setup for the $b-$quark action featured gauge links that were stout-smeared \cite{ref:stout} in all four spacetime directions with the smearing parameter $\rho = 0.125$ and $N=2$ as the number of smearing applications. Here $\nu=1$ and $c_E=c_B=c_{sw}$ would correspond to the standard Clover action. 
To tune the action \eqref{eq:action} we chose a set of observables consisting of the masses of the $B$ and $B^*$ mesons \cite{ParticleDataGroup:2024cfk}, $m_{B^{(*)}}$, and their continuum energy-momentum dispersion relations. The aim was to find the parameters that best reproduce these observables on the employed ensemble. The lattice energies were extracted from $B$ and $B^*$ two-point correlator functions using single-exponential fits in the plateau region. An example of a fit used to extract the $B^{(*)}$ ground-state masses is shown in Fig.~\ref{fig:meff_plot}. \\
\indent  The tuning strategy was executed in several distinct steps, simplified by the fact that there is a clearly ordered and naturally expected hierarchy of contributions of each parameter to a given observable. Namely, the $B$ and $B^*$ masses were by and large determined by the value of the heavy quark mass $m_Q$, while the small hyperfine mass splitting $m_{B^*} - m_{B}$ could be tuned more accurately by adjusting the Clover coefficients $c_E$ and $c_B$ without spoiling the meson masses. The parameter $\nu$ turned out to be primarily responsible for the constant $c$ appearing in the relativistic energy-momentum dispersion relations
\begin{align}\label{eq:en-mom}
    aE &= \sqrt{ (am)^2 +  a^2(c\vec p)^2 }, \hspace{1cm} \vec p = \frac{2 \pi}{aN_L} \vec n, \ \vec n \in \mathbb{Z}^3  
\end{align}
The resulting $B$, $B^*$ and $T_{bb}$ energy-momentum dispersion relations based on the tuned parameters are summarized by Fig.~\ref{fig:disprels} and Table \ref{tab:disprels}.
\begin{table}[h!]
  \centering
  \begin{tabular}{l  l l  }
    \toprule
    $h$ & $m_h$ (GeV) & $c$ \\
    \midrule
    $B$ & $5.3020(17)$& $1.020(12)$  \\
     $B^*$ & $5.3387(20)$& $1.021(14)$ \\
     \midrule
    $T_{bb}$ & $10.5765(98)$ & $1.04(11)$   \\
    \bottomrule
  \end{tabular}
  \caption{Masses and values of the speed of light for $B$, $B^*$ and  $T_{bb}$. These were determined from fits of Eq.~\eqref{eq:en-mom} to the lattice energies shown in Fig.~\ref{fig:disprels}. }
  \label{tab:disprels}
\end{table}
\supsection{Interpolators, correlators and renormnalization }
\subsection{Interpolators}
We employed a single local interpolator for each type of hadron $h = B, \ B^*,  \ \pi, \ T_{bb}$ to compute two- and three-point correlators:
\begin{gather}
    \mathcal{O}_{B}^\dagger  (x) = \bar{b}(x) \gamma_5 u(x), \nonumber \\[5pt]
    \mathcal{O}_{B^*}^\dagger  (x, i) = \bar{b}(x) \gamma_i u(x), \label{eq:interps} \\[5pt]
    \mathcal{O}_{\pi}^\dagger  (x) = \bar{d}(x) \gamma_5 u(x), \nonumber \\[5pt]
    \mathcal{O}_{T_{bb}}^\dagger  (x, i) = \epsilon^{abc} \epsilon^{ade} [ \bar{b}(x)_b C \gamma_i \bar{b}(x)_c ] [ u(x)_{d} C \gamma_5 d(x)_e ], \nonumber
\end{gather}
where $C$ labels the charge conjugation matrix. The overlap $\mathcal{Z}_0 = \langle 0 | \mathcal{O}_h^\dagger | \Omega \rangle$ of the operator to the ground state (labeled $0$)
was improved by applying the standard gauge-covariant spatial Gaussian smearing to all quark fields in \eqref{eq:interps} (see Eq. (8) in \cite{ref:tbb_leskovec}). For the light quarks the width parameter was set to $\sigma_q=4.5$ (in units of lattice spacing) with the number of smearing hits $N_q = 85$ using APE spatially-smeared gauge links with parameters $\alpha_{\mathrm{APE}} = 2.5$ and $N_g = 50$ smearing steps \cite{APE:1987ehd}. In the case of the heavy quarks we also used Gaussian smearing with width set to $\sigma_Q = 2.5$ and $N_Q = 15$ smearing applications using spatially-stout smeared gauge links with parameters $\rho = 0.125$ and $N_g = 5$ hits \cite{ref:stout}. The $h=B^*,T_{bb}$ interpolators were projected to the appropriate rows $r$ and irreducible representations~(irreps) $\Lambda$ of the subgroups of the octahedral group to account for the reduced rotational symmetry present in the finite-volume. The projected operators were constructed via linear combinations of interpolators featuring different Dirac gamma matrices $\gamma_i$
\begin{align}\label{eq:irrep_projections}
    \mathcal{O}_{h} (\vec p, \Lambda ,r) = \sum_{i=x,y,z} \mathcal{C}^{\Lambda, r}_i \mathcal{O}_{h} (\vec p, i),
\end{align}
where the subduction coefficients $\mathcal{C}^{\Lambda, r}_i$ are calculable using group-theoretic means, e.g. the projection formula (A5) given in \cite{Dudek:2010wm} with irreps defined in Appendix A of \cite{Bernard:2008ax}.
\subsection{Two-point correlators}
The energies and overlaps were calculated from the two-point correlation functions 
\begin{align}\label{eq:twopt}
    C_2(\vec q , t) = \langle \Omega | \mathcal{O}_h(\vec q, t) O_{h}^\dagger (0) | \Omega \rangle = 
     \sum_{n=0}^\infty \frac{|\mathcal Z_n (\vec q )|^2}{2 E_n(\vec q^2)} e^{-E_n(\vec q^2) t}
\end{align}
for four hadrons $h=T_{bb},B,B^*,\pi$ using the interpolators defined in Eq.~\eqref{eq:interps}. The momenta $\vec q$ listed in Table \ref{tab:momenta} and their permutations were used to extract the ground-state energies $E_0 (\vec q^2)$ and the overlap factors $\mathcal{Z}_0 (\vec q) = \langle 0 | \mathcal{O}_h^{\dagger} (\vec q) | \Omega \rangle$. The fitting was performed in the plateau region of the correlators using a single-exponential model  $C_2(\vec q , t)=Ae^{-Et}.$ 
We evaluated the two-point correlators for each hadron at eleven different spatial positions at the source 
and averaged them before fitting  in order to decrease the statistical uncertainty.
\begin{table}[t!]
  \centering
  \begin{tabular}{c}
    \toprule
    $\frac{L}{2\pi}\vec q \equiv \vec n \in \mathbb{Z}^3$ \\
    \midrule
    $(0,0,0)$  \\
    $(0,0,1)$  \\
    $(1,1,0)$    \\
    $(1,1,1)$   \\
    $(0,0,2)$ \\
    $(0,0,3)$ \\
    \bottomrule
  \end{tabular}
  \caption{List of momenta $\vec q$ to which the two- and three-point correlation functions were projected at the sink and the current, respectively. The permutations of each vector $\vec n$ are not displayed here for brevity, but were also used in the lattice computations.}
  \label{tab:momenta}
\end{table}
\subsection{Three-point correlators}\label{subsec:setup}
The three point correlators $C_3^\mu$, defined as
\begin{align}\label{eq:threept_corr2}
    \mathcal{C}_{3}^\mu &(\vec p_2, \vec q, T; t) = \langle \Omega | \mathcal{O}_{h} ( \vec p_2, T) \hat \jmath_{EM}^\mu (\vec q, t) \mathcal{O}^\dagger_{h} (  0) | \Omega \rangle = 
     \sum_{n,m = 0}^\infty \frac{\mathcal{Z}^{f*}_n \mathcal{Z}^i_m}{(2E^f_n)(2E^i_m)}  \mathcal{M}_{nm}^\mu \ e^{-E^f_n (T-t)} e^{-E^i_mt}, 
\end{align}
were computed using the renormalized electromagnetic (EM) current
\begin{align}\label{eq:current2}
    \hat \jmath^\mu_{EM}  = \sum_{q \in \{ u,d,b \}} Z_q^V \cdot  e_q \bar q \gamma^\mu q.
\end{align}
Here the sum goes over quark fields $q$ and their electric charges $e_q$, while the determination of the renormalization constants $Z_q^V$ is discussed in the next subsection.\\
\indent The connected diagrams were constructed in the standard way from sequential and forward propagators. 
In this approach, the sink momentum $\vec p_2$ and the source-sink separation $T$ are kept fixed. We set $\vec p_2= \vec 0$ and realized the source-sink separations: $\tfrac{T}{a} = 12,15,18,22$. The set of momenta $\vec q$ used in projecting the current to definite momentum is listed in Table \ref{tab:momenta}. Consequently, due to momentum conservation, the available three-point correlator data enabled us to access ground-state matrix elements with momenta $\langle h (\vec p_2 = \vec 0) | \hat \jmath_{EM}^\mu |h (\vec p_1  = -\vec q) \rangle$.\\
\subsection{Renormalization of the electromagnetic current}\label{subsec:renorm}
The continuum and infinite-volume normalization of the elastic charge form factor for any particle is well-known
\begin{align}\label{eq:normalization}
    F_C(Q^2=0) = Z,
\end{align}
where $Z$ is charge of the particle in units of the elementary charge $e_0$. In total, two renormalization factors had to be determined: $Z^V_b$ and $Z^V_{u/d}$, corresponding to the renormalization of heavy- and light-quark EM currents, respectively, defined as
\begin{gather}
    \hat \jmath^\mu_{u/d} = Z^V_{u/d} \cdot \left( \tfrac{2}{3} \bar u \gamma^\mu u - \tfrac{1}{3} \bar d \gamma^\mu d \right), \hspace{0.4cm} \hat \jmath_{b}^\mu = Z^V_b \cdot \left( -\tfrac{1}{3} \bar b \gamma^\mu b\right), \nonumber \\[10pt]  \hat \jmath^\mu_{EM} \equiv  \hat \jmath^\mu_{u/d} + \hat \jmath_{b}^\mu . \label{eq:curre}
\end{gather}
 These factors were calculated by imposing the condition \eqref{eq:normalization} on the renormalized charge form factors of the $b$ and $\bar u$ quarks within the $T_{bb}$
\begin{gather}
    F_C^{b,T_{bb}}(0) \equiv -\frac{2}{3}, \hspace{0.4cm} F_{C}^{\bar u, T_{bb}}(0) \equiv -\frac{2}{3}, 
    \label{eq:normffs}
\end{gather}
thereby fixing values of both factors
\begin{gather}
    Z^V_{b} = 3.97(11), \hspace{0.4cm} Z_{u/d}^V = 0.725(16).
    \label{eq:renormcoeffs}
\end{gather}
The same set of renormalization factors was applied to recover the infinite-volume continuum normalizations of $B,B^*$ and pion form factors
\begin{gather}
    F_C^{B}(0) = -1.003(24), \hspace{0.4cm} F_C^{B^*}(0) = -1.014(25), \hspace{0.4cm} F_{C}^\pi (0) = -0.957(24). 
    \label{eq:bbstarFF}
\end{gather}
We remark that as we use an anisotropic action for the heavy quark, we
should, in principle, allow for different renormalization factors for
the temporal and spatial components of the heavy quark vector current
$\bar{b}\gamma^\mu b$. However, we have verified that the results for the form factor
$F_1\approx F_C$ extracted from the temporal and spatial components of
the corresponding three-point functions agree well within the statistical
uncertainties. Considering the systematics inherent in the
calculation, it is sufficient to set the temporal and spatial
renormalisation factors to be equal.
\supsection{Fitting three-point correlator data}\label{sec:dataprocessing}
The three-point correlators were further processed by constructing two types of ratio quantities, $R_3^\mu$ and $R_{3,symm}^\mu$, with the first one defined in an identical manner to eq. (28) of \cite{PhysRevLett.134.161901} and the other featuring built-in $t \rightarrow T - t$ symmetry 
\begin{gather}
    R_3^\mu (\vec p_2, \vec q, T; t)  = \frac{(2E^f_0)(2E^i_0)}{\mathcal{Z}^{f*}_0 \mathcal{Z}^i_0}e^{E^f_0(T-t)} e^{E^i_0 t}C_3^\mu (\vec p_2, \vec q, T; t), \label{eq:ratio3pt0}\\[10pt]
    R_{3,symm}^\mu (\vec p_2, \vec q, T; t)  = \mathrm{sgn} \left(R_3^\mu (\vec p_2, \vec q, T; t) \right) \cdot \sqrt{|R_3^\mu (\vec p_2, \vec q, T; t) \cdot R_3^\mu (\vec p_2, \vec q, T; T-t)|} .\label{eq:ratio3pt1}
\end{gather}
Using the decomposition in eq.~\eqref{eq:threept_corr2} it can be easily shown that both ratios equal the ground-state matrix element, up to excited-state contamination. Note that, in addition to the dependence on the momenta $\vec p_2$ and  $\vec p_1 = \vec p_2 - \vec q$ (which is kept explicit throughout the Letter and this Supplemental Material), the ratios also depend on the rows $r_2, r_1$ of the irreps $\Lambda_2, \Lambda_1$ to which they are projected at the sink and the source, respectively. To simplify the notation, these additional indices are not shown unless needed.\\
\indent Prior to fitting,  we applied a weighted averaging procedure to the ratios, in order to decrease noise in the lattice data. This was based on the fact that the form factors contained in the matrix elements depend merely on the Lorentz-scalar $Q^2$, carrying no information about the directions of the source and sink three-momenta. As discussed previously in Subsection \ref{subsec:setup}, the momentum at the sink was always set to zero in our setup, meaning that all matrix elements with source momenta lying on the same shell ($|\vec p_1|^2 = const.$) yield the form factors at a constant value of $Q^2$. This enabled us to define the weighted direction-averages of the ratios $\bar{R}_{3(,symm)}^\mu$
\begin{gather}
    \bar{R}_{3(,symm)}^\mu (Q^2, T; t) = \frac{1}{N_{tot}} \sum_{\substack{\Lambda_1, r_1,\vec p_1, \Lambda_2, r_2 \\ |\vec p_1|^2=const.}} w_{\Lambda_1, r_1, \vec p_1}^{\Lambda_2, r_2} \cdot R_{3(,symm)}^\mu (\vec p_2 = \vec 0, \Lambda_2, r_2; \Lambda_1, r_1, \vec p_1 \equiv -\vec q; T; t) \propto F_i(Q^2), \label{eq:avgratio}
\end{gather}
where all irrep and momenta dependence is now kept explicit for clarity. The sum runs over all possible combinations that yield a constant $Q^2$, while the weights $w$ are explicitly provided in Subsection \ref{subsec:ffext}. We take the ratios in \eqref{eq:avgratio} as the final lattice-derived quantities to which we perform fits.\\
\subsection{Fit models}
\indent To extract the matrix elements $\mathcal M$ from direction-averaged ratios, we employed three basic models: a constant model \texttt{const}, assuming no contamination is present in the signal, that is fitted straightforwardly to both unsymmetrized and symmetrized ratios 
\begin{align}\label{eq:constfit}
    f_{\texttt{const}}(t) = \mathcal M,
\end{align}
or a \texttt{2exp} model, incorporating first-excited states at the source and the sink, applicable only to unsymmetrized $R_3^\mu$
\begin{align}\label{eq:2expfit}
    f_{\texttt{2exp}} (t) = \mathcal M+ \alpha e^{-\Delta E_1t} + \beta e^{-\Delta E_2 (T-t)},
\end{align}
where $\mathcal{M}, \alpha, \beta, \Delta E_1, \Delta E_2$ all represent fit parameters. The first-excited-state model \texttt{symm2exp} used for fitting the symmetrized ratio \eqref{eq:ratio3pt1} is 
\begin{align}\label{eq:symm2expfit}
    f_{\texttt{symm2exp}} (t) = \mathcal M+ \alpha  \left( e^{-\Delta Et} +  e^{-\Delta E (T - t)} \right),
\end{align}
with $\mathcal M, \alpha, \Delta E$ being the fit parameters. All fits were performed with a fully correlated covariance matrix simultaneously to four sets of data with different source-sink separations $T$. Plots showing some of the typical fits for $T_{bb}, B, B^*$ can be seen in Section \ref{sec:Plots featuring fits of the matrix elements}.
\subsection{Form factor extraction}\label{subsec:ffext}
In the case of pseudoscalar mesons $h=B, \pi$ the procedure for obtaining the single charge form factor from the data was straightforward. It involved dividing out the kinematic prefactor $(p_1+p_2)^\mu$ from the fitted matrix element that appears in the decomposition
\begin{align}
    \mathcal{M}_{EM}^\mu = (p_1 + p_2)^\mu F_C(Q^2) \xrightarrow{\mu = 0} (E_1 + E_2) F_C(Q^2).
\end{align}
The weights needed for direction-averaged ratios \eqref{eq:avgratio} at $\mu = 0$ are trivial, $w_i = 1$. \\
\indent In contrast, the analogous procedure was somewhat more involved for the $T_{bb}$ and $B^*$. Both hadrons share an identical EM form factor decomposition
\begin{align}
    \mathcal{M}_{EM}^\mu =&-(p_1 + p_2)^\mu (\varepsilon_2^*  \cdot \varepsilon_1  ) F_1 (Q^2 ) - \label{eq:decomp_vax} \\[5pt]
    &- [(\varepsilon_2^*  \cdot q)\varepsilon_1^\mu - (\varepsilon_1  \cdot q)\varepsilon_2^{*\mu}] F_2(Q^2) + \nonumber \\[5pt]
    &+ \frac{(\varepsilon_2^* \cdot q) (\varepsilon_1 \cdot q)}{2m^2} (p_1 + p_2)^\mu F_3(Q^2) \nonumber = \\
    &\equiv a_1 F_1(Q^2) + a_2 F_2(Q^2) + a_3 F(Q^2), \label{eq:decomp_vax2}
\end{align}
so we only discuss $T_{bb}$ here. The strategy we adopted involved extracting form factors $F_1, F_2$ and $F_3$ in steps. The first two form factors were obtained by considering only particular momenta and irrep configurations in which a single coefficient $a_1$ or $a_2$ in \eqref{eq:decomp_vax2} is non-vanishing, requiring only to divide out the kinematic prefactor $a_i$ from the fitted matrix element as in the pseudoscalar case. The third form factor, contrary to others, could not be directly obtained in isolation and was calculated using the fits and the already known $F_1, F_2$. \\
\indent For form factors $F_{1(2)}$, the weights in \eqref{eq:avgratio} were accordingly chosen to be nonzero only for the ratios that feature a single nonvanishing coefficient $a_{1(2)}$ in \eqref{eq:decomp_vax}. For the $F_3$, we averaged over all momenta and irrep configurations that give identical, but nonzero $a_1, a_2$ and $a_3$. All such configurations are collected in comprehensive tables in Section \ref{sec:Parameters and covariances of $z-$expansion fits to the form factors}. 

In the final step we converted the original form factor basis $F_{1-3}$ into the charge, magnetic dipole and electric quadrupole form factors $F_C, F_M, F_{\cal Q}$ using a linear transformation
\begin{align}\label{eq:multipoleFF}
    \begin{pmatrix}
    F_C (Q^2) \\[4pt]
    F_M (Q^2) \\[4pt]
    F_{\cal{Q}} (Q^2)
    \end{pmatrix}
    =
    \begin{pmatrix}
    1+\frac{2}{3}\eta & - \frac{2}{3}\eta & \frac{2}{3}\eta(1+\eta) \\[4pt]
    0 & 1 & 0 \\[4pt]
    1 & -1 & (1 + \eta)
    \end{pmatrix}
    \begin{pmatrix}
    F_1 (Q^2) \\[4pt]
    F_2 (Q^2) \\[4pt]
    F_3 (Q^2)
    \end{pmatrix} ,
\end{align}
with $\eta = \frac{Q^2}{4m^2}$.
\supsection{Parameters and covariances of $z-$expansion fits to the form factors}
The general parametrization of a form factor $F(Q^2)$ for $Q^2 \equiv -q^2 \geq 0$ is given by \cite{Boyd:1994tt,Boyd:1997qw,Bourrely:2008za}
\begin{align}
\label{z-exp}
    F(Q^2) = \frac{1}{1 + \frac{Q^2}{m_r^2}}\sum_n a_n z^n(Q^2; t_+, t_0),
\end{align}
with the $z-$variable defined as
\begin{align}
    z(Q^2; t_+, t_0) = \frac{\sqrt{t_+ + Q^2} - \sqrt{t_+ - t_0}}{\sqrt{t_+ + Q^2} + \sqrt{t_+ - t_0}} .
\end{align}
If the form factor has an additional constraint at $Q^2=0$, i.e. $F(0) = F_0$, then the zeroth-order parameter $a_0$ is fixed and obeys the relation
\begin{align}\label{eq:zexp_constraint}
    a_0 = F_0 - \sum_{n \neq 0}^{n_{max}} a_n z^n(0; t_+, t_0) .
\end{align}
The values of the resonance masses $m_r$, multi-particle thresholds $t_+$ and $t_0$ used in $z-$expansion parametrizations are listed in Table \ref{tab:zexp_params}. Numerical values of the fit parameters $a_n$, as well as their correlation matrices, are shown in Tables \ref{tab:FC-others}-\ref{tab:FQ-Tbb}. They were used to generate Figs. 2 and 3 in the Letter. The charge form factors of the $B,B^*$ and pion serve as a reference point for the form factors of the $T_{bb}$. The results for the form factors $F_C$, $F_M$ and $F_{\cal Q}$ of the $T_{bb}$ are collected in Tables \ref{tab:FC-Tbb},  \ref{tab:FM-Tbb} and  \ref{tab:FQ-Tbb}, respectively: these provide the form factors based on the total EM current, as well as separate contributions from the heavy and light currents  (\ref{eq:curre}). Note that the zeroth-order parameter $a_0$ is absent in the charge form factors $F_C$ because it is constrained by eq.~\eqref{eq:zexp_constraint} and $F_C(0) = -1$. This condition holds exactly for the $T_{bb}$ as covered in Subsection \ref{subsec:renorm} due to our choice of calculating the EM current renormalization factors based on $T_{bb}$ matrix elements. For the $B,B^*$ and  $\pi$ it constitutes a very good approximation, as can be seen from eq.~\eqref{eq:bbstarFF}.
\begin{table}[htb]
\centering
\begin{minipage}{0.45\linewidth}
  \centering
\begin{tabular}{l c c}
    \toprule
    $h$ & $m_r \ (\mathrm{GeV})$ & $t_+ \ (\mathrm{GeV}^2)$  \\
    \midrule
    $T_{bb}$ & $m_\omega = 0.7877$  & $(3m_\pi)^2 = 0.7521(50)$   \\
     $B, B^*$ & $m_\rho = 0.7655$  & $(2m_\pi)^2 = 0.3343(22)$   \\
     $\pi$ & $m_\rho $  & $(2m_\pi)^2  $  \\
    \bottomrule
  \end{tabular}
\end{minipage}
\hfill
\begin{minipage}{0.45\linewidth}
  \centering
  \begin{tabular}{l c c}
    \toprule
    $h$ & form factor  & $t_0$  \\
    \midrule
    $B, B^*$ & $F_C$  & $-10 \cdot (2m_\pi)^2$   \\
    $T_{bb}$ & $F_M, F_M^{[bb]}$  & $-15 \cdot  (3m_\pi)^2$   \\
    \midrule
    All others & & 0 \\
    \bottomrule
  \end{tabular}
\end{minipage}
\caption{Numerical values of resonance masses $m_r$,  multi-particle thresholds $t_+$  and parameter $t_0$ used in the $z-$expansions. The same values for $m_r$ and $t_+$ were utilized for all form factors of a given hadron $h$. The value $t_0=0$ is used, except in the cases indicated in the right table.}
  \label{tab:zexp_params}
\end{table}
\begin{table}[ht!]
\begin{center}
    \begin{minipage}{0.32\linewidth}
          \centering
          \begin{tabular}{l c c c}
          $F_C^B$ & Value & \multicolumn{2}{c}{Corr. matrix} \\
          \toprule
             $a_1$ & $-5.72(45)$ & $1.0$ & $0.96$  \\
             $a_2$ & $-5.86(59)$ & $0.96$ & $1.0$ \\
            \bottomrule
          \end{tabular}
    \end{minipage} \hfill
    \begin{minipage}{0.32\linewidth}
    \centering
          \begin{tabular}{l c c c}
          $F_C^{B^*}$ & Value & \multicolumn{2}{c}{Corr. matrix} \\
          \toprule
             $a_1$ & $-5.17(49)$ & $1.0$ & $0.96$  \\
             $a_2$ & $-5.42(65)$ & $0.96$ & $1.0$ \\
            \bottomrule
          \end{tabular}
    \end{minipage}
    \begin{minipage}{0.32\linewidth}
    \centering
         \begin{tabular}{l c }
  $F_C^{\pi}$ & Value  \\
    \toprule
     $a_1$ & $0.11(15)$   \\
    \bottomrule
  \end{tabular}          
    \end{minipage}
     \caption{Parameters of the fit (\ref{z-exp}) for the electric charge form factors $F_C$  of hadrons $h=B,B^*,\pi$, which serve as a reference. 
     A second-order fit is used for the $B$ and $B^*$ form factors, while a first-order expansion is used for the pion.}\label{tab:FC-others}
\end{center}
\end{table}

\begin{table}[ht!]
\begin{center}
\begin{minipage}{0.32\textwidth}
    \centering
          \begin{tabular}{l c c c}
          $F_C^{T_{bb}}$ & Value & \multicolumn{2}{c}{Corr. matrix} \\
          \toprule
             $a_1$ & $-1.63(42)$ & $1.0$ & $-0.9$  \\
             $a_2$ & $-8.6(1.5)$ & $-0.9$ & $1.0$ \\
            \bottomrule
          \end{tabular}
          \end{minipage} \hfill
    \begin{minipage}{0.32\textwidth}
          \centering
          \begin{tabular}{l c c c}
          $F^{[bb]}_C$ & Value & \multicolumn{2}{c}{Corr. matrix} \\
          \toprule
             $a_1$ & $-2.86(31)$ & $1.0$ & $-0.85$  \\
             $a_2$ & $-7.5(1.1)$ & $-0.85$ & $1.0$ \\
            \bottomrule
          \end{tabular}
    \end{minipage} \hfill
    \begin{minipage}{0.32\textwidth}
    \centering
          \begin{tabular}{l c c c}
          $F^{[\bar u\bar d]}_C$ & Value & \multicolumn{2}{c}{Corr. matrix} \\
          \toprule
             $a_1$ & $1.76(18)$ & $1.0$ & $-0.95$  \\
             $a_2$ & $-2.52(70)$ & $-0.95$ & $1.0$ \\
            \bottomrule
          \end{tabular}
    \end{minipage} \hfill
    \caption{Parameters of the fit (\ref{z-exp}) for the charge form factors $F_C$ of $T_{bb}$. Left: The charge form factor of the $T_{bb} $. Middle: heavy diquark charge form factor of the $T_{bb}$. Right: the light antidiquark charge form factor of the $T_{bb}$. All fits of $F_C$ are done using the second order in the $z-$expansion.}\label{tab:FC-Tbb}
\end{center}
\end{table}

\begin{table}[ht!]
\begin{center}
    \begin{minipage}{0.32\textwidth}
        \centering
              \begin{tabular}{l c c c c}
              $F_M^{T_{bb}}$ & Value & \multicolumn{2}{c}{Corr. matrix} \\
              \toprule
                 $a_0$ & $17.56(98)$ & $1.0$ & $0.99$ & $0.98$  \\
                 $a_1$ & $37.6(3.9)$ & $0.99$ & $1.0$ & $0.99$  \\
                 $a_2$ & $19.1(3.9)$ & $0.98$ & $0.99$ & $1.0$ \\
                \bottomrule
              \end{tabular}
        \end{minipage} \hfill
    \begin{minipage}{0.32\textwidth}
    \centering
          \begin{tabular}{l c c c c}
          $F_M^{[bb]}$ & Value & \multicolumn{2}{c}{Corr. matrix} \\
          \toprule
                 $a_0$ & $18.41(70)$ & $1.0$ & $0.99$ & $0.97$  \\
                 $a_1$ & $41.1(2.7)$ & $0.99$ & $1.0$ & $0.99$  \\
                 $a_2$ & $22.5(2.6)$ & $0.97$ & $0.99$ & $1.0$ \\
                \bottomrule
          \end{tabular}
    \end{minipage} \hfill
    \begin{minipage}{0.32\textwidth}
    \centering
          \begin{tabular}{l c c c c}
          $F_{M}^{[\bar u \bar d]}$ & Value & \multicolumn{2}{c}{Corr. matrix} \\
          \toprule
                 $a_0$ & $0.02(18)$ & $1.0$ & $-0.99$ & $0.99$  \\
                 $a_1$ & $0.1(2.1)$ & $-0.99$ & $1.0$ & $-0.99$  \\
                 $a_2$ & $-0.3(4.8)$ & $0.99$ & $-0.99$ & $1.0$ \\
                \bottomrule
          \end{tabular}
    \end{minipage} \hfill
     \caption{ Parameters of  the fit (\ref{z-exp}) for the magnetic dipole form factors $F_M$ related to $T_{bb}$. Left: The magnetic dipole form factor of the $T_{bb}$ Middle: heavy diquark magnetic dipole form factor of the $T_{bb}$. Right: the light antidiquark magnetic dipole form factor of the $T_{bb}$. All fits of $F_M$ are done to the second order in $z-$ expansion. }\label{tab:FM-Tbb}
\end{center}
\end{table}

\begin{table}[ht!]
\begin{center}
    \begin{minipage}{0.25\textwidth}
        \centering
              \begin{tabular}{l c }
              $F_{\cal Q}^{T_{bb}}$ & Value \\
              \toprule
                 $a_0$ & $2(76)$  \\
                \bottomrule
              \end{tabular}
        \end{minipage} \hfill
    \begin{minipage}{0.25\textwidth}
    \centering
          \begin{tabular}{l c }
          $F^{[bb]}_{\cal Q}$ & Value  \\
          \toprule
                 $a_0$ & $1(75)$   \\
                \bottomrule
          \end{tabular}
    \end{minipage} \hfill
    \begin{minipage}{0.32\textwidth}
    \centering
          \begin{tabular}{l c c c}
          $F^{[\bar u\bar d]}_{\cal Q}$ & Value & \multicolumn{2}{c}{Corr. matrix} \\
          \toprule
                 $a_0$ & $-0.18(85)$ & $1.0$ & $-0.95$ \\
                 $a_1$ & $3.5(4.1)$ & $-0.95$ & $1.0$ \\
                \bottomrule
          \end{tabular}
    \end{minipage} \hfill
    \caption{ Parameters of the $z-$expansion fit (\ref{z-exp}) for the  electric quadrupole form factors $F_{\cal Q}$ related to $T_{bb}$. Left: Zeroth-order fit for the quadrupole form factor of the $T_{bb}$. Middle: Zeroth-order fit for the heavy diquark quadrupole form factor of the $T_{bb}$. Right: First-order fit for the light antidiquark quadrupole form factor of the $T_{bb}$. }\label{tab:FQ-Tbb}
    \end{center}
\end{table}

\newpage

\supsection{Charge density of the $T_{bb}$ in position-space}

\begin{figure*}[!t]
    \begin{overpic}[width=\textwidth]{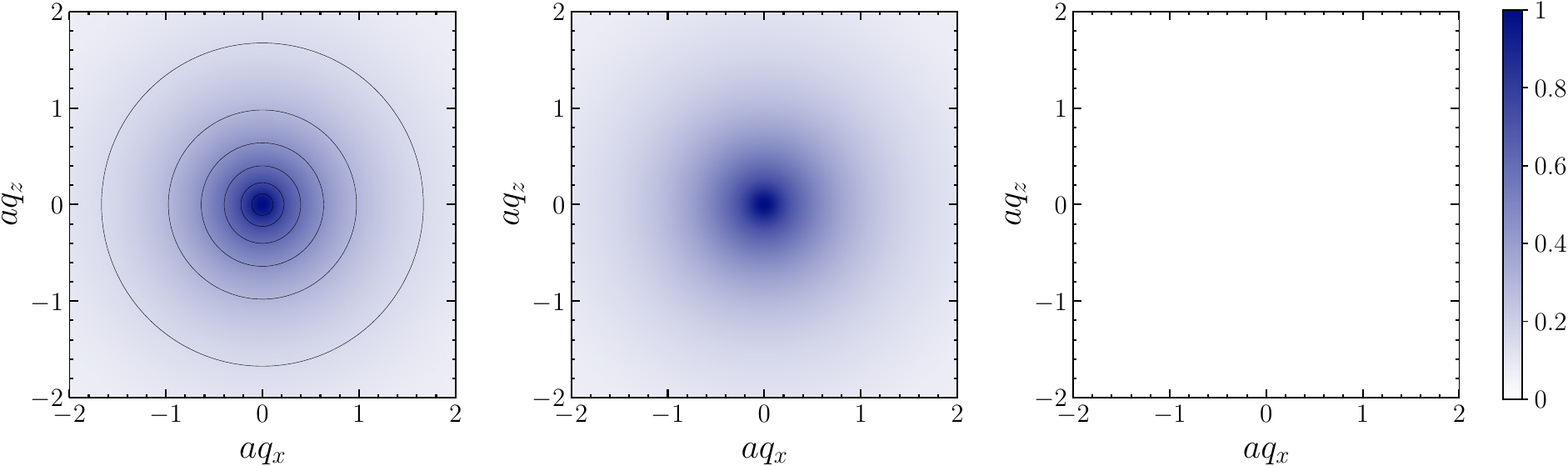}
        \put(0,31){\textbf{a)} $|G_{0} (|\vec q|^2) Y_{l0}(\Omega_q) + |\vec q|^2 G_{2} (|\vec q|^2) Y_{20}(\Omega_q)|$}
        \put(40,31){\textbf{b)} $|G_{0} (|\vec q|^2) Y_{l0}(\Omega_q)|$}
        \put(71,31){\textbf{c)} $||\vec q|^2 G_{2} (|\vec q|^2) Y_{l0}(\Omega_q)|$}
    \end{overpic}
    \caption{\textbf{a)} Electric charge density $\rho(\vec q)$ in momentum space at $aq_y=0$, based on eq.~\eqref{eq:multipole2} with contour lines shown. \textbf{b)} and \textbf{c)} Separate contributions of the two multipoles, $l=0$ and $2$, that are accessible from the form factors $G_{l0}$ which are related to electric form factor $F_l$ via eq. (\ref{eq:FFtransformation}). Index $m\!=\!0$ in $G_{lm}$ is omitted for brevity. } 
    \label{fig:rho_q}
\end{figure*}

\indent The momentum-space charge density $\rho(\vec q)$ is given by the position-space charge distribution via
\begin{align}\label{eq:chdensityq}
    \rho (\vec q) \equiv \int \mathrm{d}^3r \ e^{-i \vec q \cdot \vec r} \ \rho(\vec r) \rightarrow \rho(\vec r) = \int \frac{\mathrm{d}^3q}{(2\pi)^3} \ e^{i\vec q \cdot \vec r} \ \rho(\vec q) .
\end{align}
It can be decomposed into an infinite tower of multipole form factors $G_{lm}$
\begin{align}\label{eq:multipole}
    \rho (\vec q) = \sum_{l=0}^\infty \sum_{m=-l}^l |\vec q|^l G_{lm} (|\vec q|^2) Y_{lm}(\Omega_q),
\end{align}
which further simplifies in the case of cylindrical symmetry along the quantization axis of the system
\begin{align}\label{eq:multipole2}
    \rho(\vec q) = \sum_{l=0}^\infty |\vec q|^l G_{l0}(|\vec q|^2) Y_{l0}(\Omega_q) .
\end{align}
The form factors $G_{l}\equiv G_{l0}$ can be shown to be directly related to the form factors we extract from the matrix elements $F_l$ in the Breit frame\footnote{Note that here we switch form factor labels $F_l \leftrightarrow G_l$ and use them in precisely the opposite way compared to \cite{Lorce:2009br}.} \cite{Lorce:2009br}
\begin{align}\label{eq:FFtransformation}
    F_{l}(|\vec q|^2) = (-i)^l \frac{(2l-1)!!}{l!} (2m)^l \sqrt{\frac{2l+1}{4\pi}} G_{l0}(|\vec q|^2) ,
\end{align}
where, more specifically, $F_{l=0}\equiv F_C$ and $F_{l=2}\equiv F_{\cal Q}$  denote the charge and quadrupole form factors used in the main text.\\ 
\indent The electric charge density $\rho(\vec q)$ is shown decomposed into the lowest two multipole contributions $l=0,2$ in Fig.~\ref{fig:rho_q}. The relative monopole contribution obtained from $F_C$ is found to be much larger than the quadropole contribution related to $F_{\cal Q}$. Likewise, the position-space charge density is therefore also dominated by the monopole term $\rho(r)$, which is shown in Fig. 2\textbf{b} of the main Letter.\\  
\indent Note that, while the form factors are generally functions of the full Lorentz scalar $Q^2 = -q^2 = -(p_2 - p_1)^2$, this quantity reduces to $|\vec q|^2 = (\vec p_2 - \vec p_1)^2$ that only depends on the three-vector $\vec q$ in the Breit frame. Our computational setup leads us to matrix elements $\langle h(\vec p_2= \vec 0) | \hat \jmath_{EM}^\mu | h(\vec p_1) \rangle$ that are generally not in the Breit frame. However, we have checked that at the values of the full momentum transfer $Q^2$ where we have data, the timelike component $-(E_2 - E_1)^2$ contributes negligibly, making $Q^2 \approx |\vec q|^2 = (\vec p_2 - \vec p_1)^2$ a very good approximation for $T_{bb}$, $B$ and $B^*$. Consequently, eq.~\eqref{eq:FFtransformation} can be used to reconstruct the momentum- and position-space charge distributions using the available form factors. \\

\FloatBarrier
\clearpage
\supsection{Plots featuring fits of the matrix elements}\label{sec:fitplots}

\begin{figure*}
    \includegraphics[width=\textwidth]{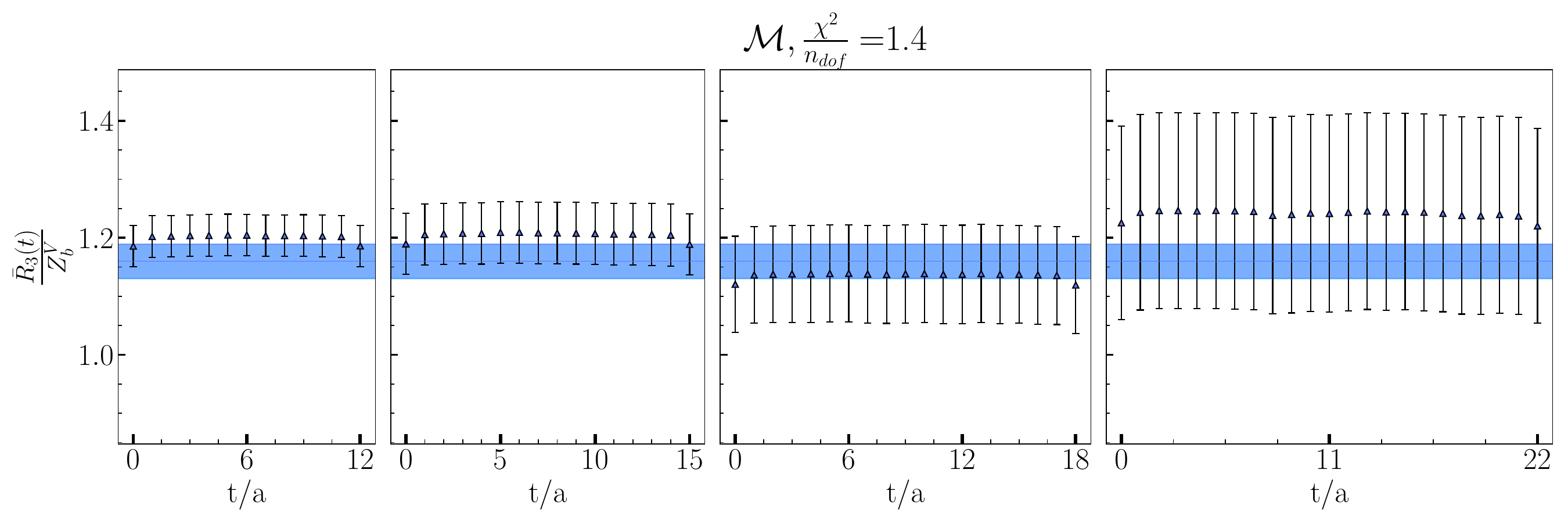}
    \caption{Depiction of the $T_{bb}$ matrix element featuring the bare heavy-quark current $\tfrac{\hat \jmath_b}{Z^V_b}$ that was used to non-perturbatively determine the renormalization factor $Z^V_b$. The \texttt{const} model \eqref{eq:constfit} was used to fit the matrix element due to absence of excited-state contamination. Discrete markers show the direction-averaged ratio \eqref{eq:avgratio}. The value of the fitted matrix element is shown with a blue line and errorbands.}
    \label{fig:ME1}
\end{figure*}

\begin{figure*}
    \includegraphics[width=\textwidth]{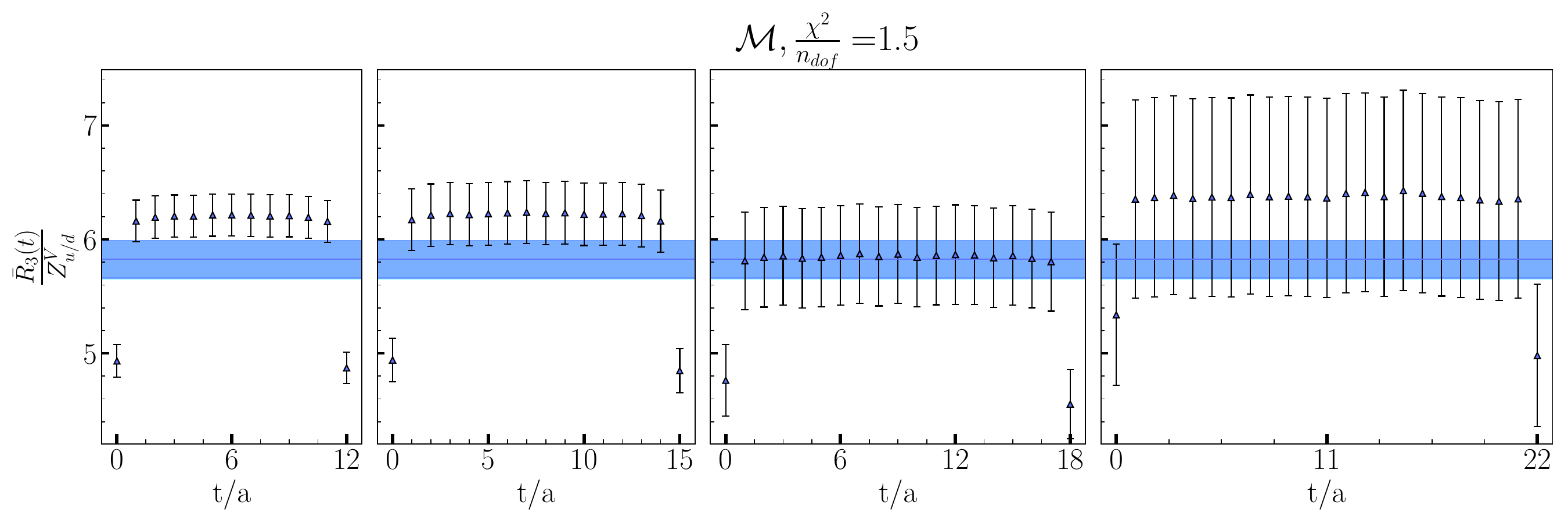}
    \caption{Depiction of the $T_{bb}$ matrix element featuring the bare $u-$quark current $\tfrac{\hat \jmath_u}{Z^V_{u/d}}$ that was used to non-perturbatively determine the renormalization factor $Z^V_{u/d}$. Other elements are the same as in Fig.~\ref{fig:ME1}.}
    \label{fig:ME2}
\end{figure*}

This section shows plots of some of the fits that were done to extract matrix elements of the $T_{bb}, B$ and $B^*$. In particular, Figs.~\ref{fig:ME1} and \ref{fig:ME2} show lattice matrix elements with yet-unrenormalized EM currents from which the factors $Z^V_{u/d}$ and $Z^V_{b}$ were determined. All subsequent figures show fully renormalized matrix elements and fits.

\FloatBarrier
\vfill

\begin{figure}
    \includegraphics[width=0.93\textwidth]{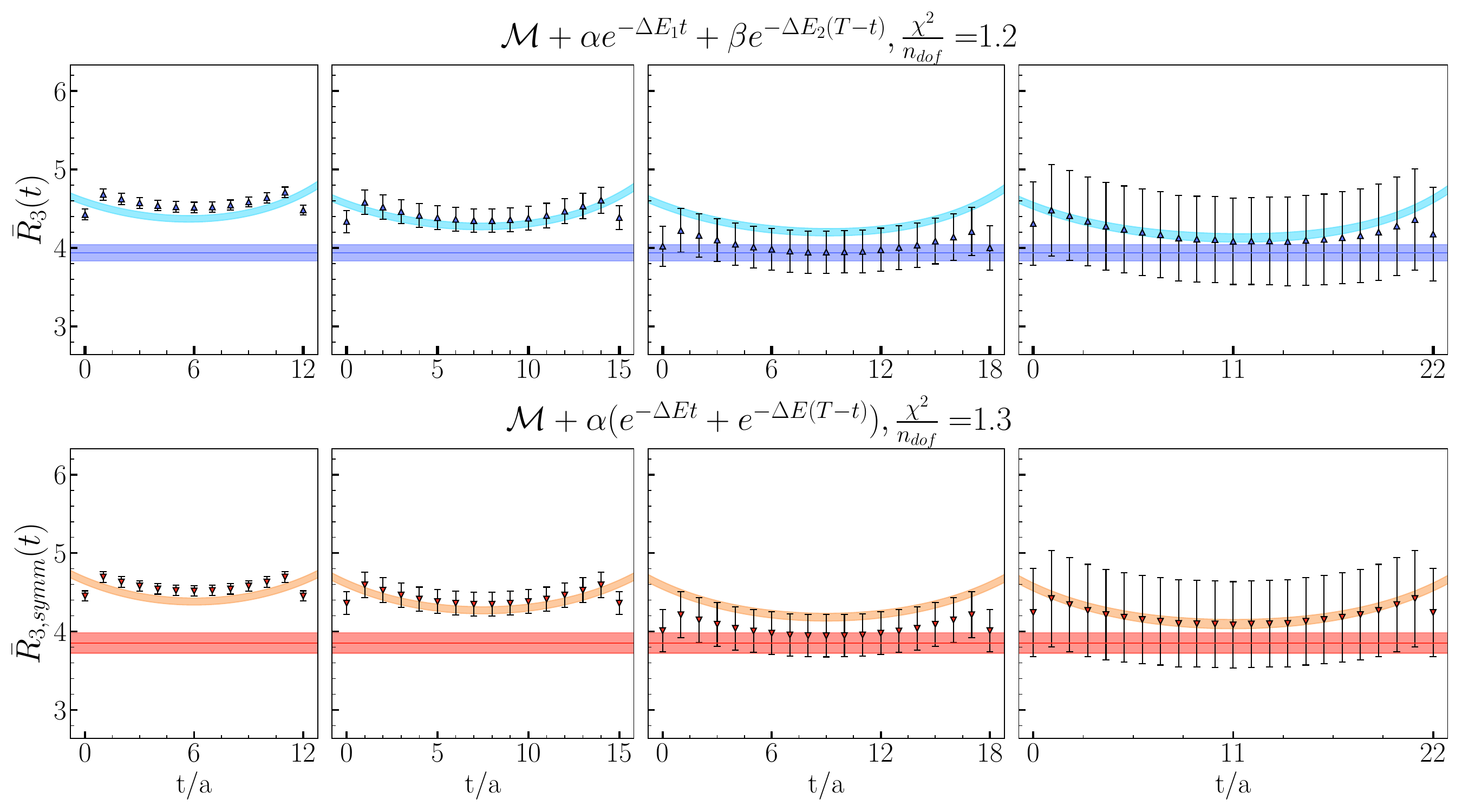}
    \caption{$T_{bb}$ matrix element featuring the full renormalized EM current used to extract the first form factor $F_1$ on the $\frac{L}{2\pi}\vec p_1 = (1,1,1)$ momentum shell. Upper and lower rows of subfigures show the unsymmetrized direction-averaged ratio $\bar{R}_3^{\mu = 0}$ and the $t \rightarrow T-t$ symmetrized ratio $\bar{R}_{3,symm}^{\mu = 0}$, respectively. The \texttt{2exp} \eqref{eq:2expfit} and \texttt{symm2exp} \eqref{eq:symm2expfit} models were used to fit both matrix elements, as indicated at the top of both subplots. Discrete markers show lattice data, curved bands show the fit function and the straight line with errorbands shows the extracted matrix element.}
    \label{fig:ME3}
\end{figure}
\vfill

\begin{figure*}
    \includegraphics[width=0.93\textwidth]{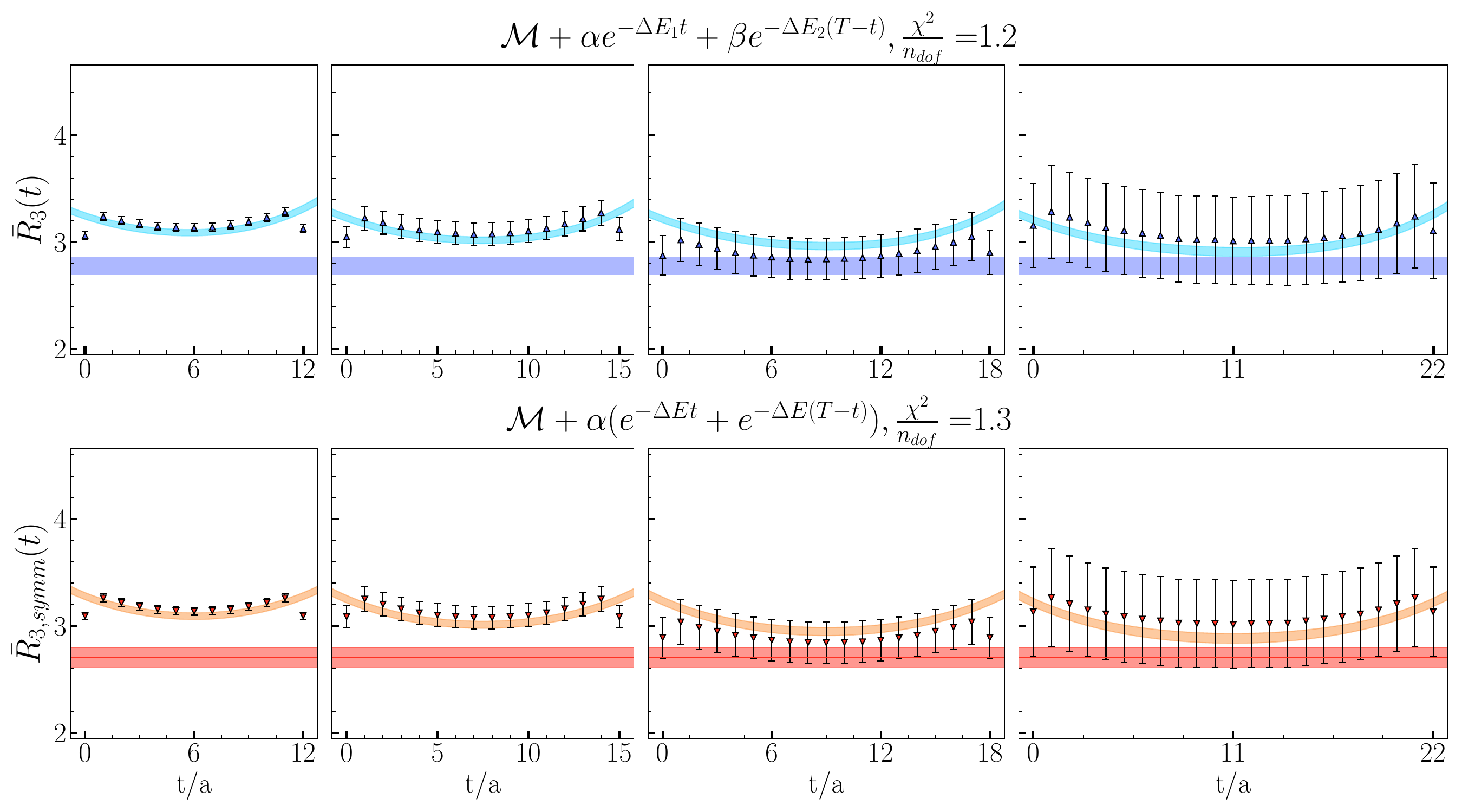}
    \caption{$T_{bb}$ matrix element featuring the full renormalized EM current used to extract the third form factor $F_3$ on the $\frac{L}{2\pi}\vec p_1 = (1,1,1)$ momentum shell. Other elements are the same as in Fig.~\ref{fig:ME3}.}
    \label{fig:ME4}
\end{figure*}

\begin{figure*}
    \includegraphics[width=0.93\textwidth]{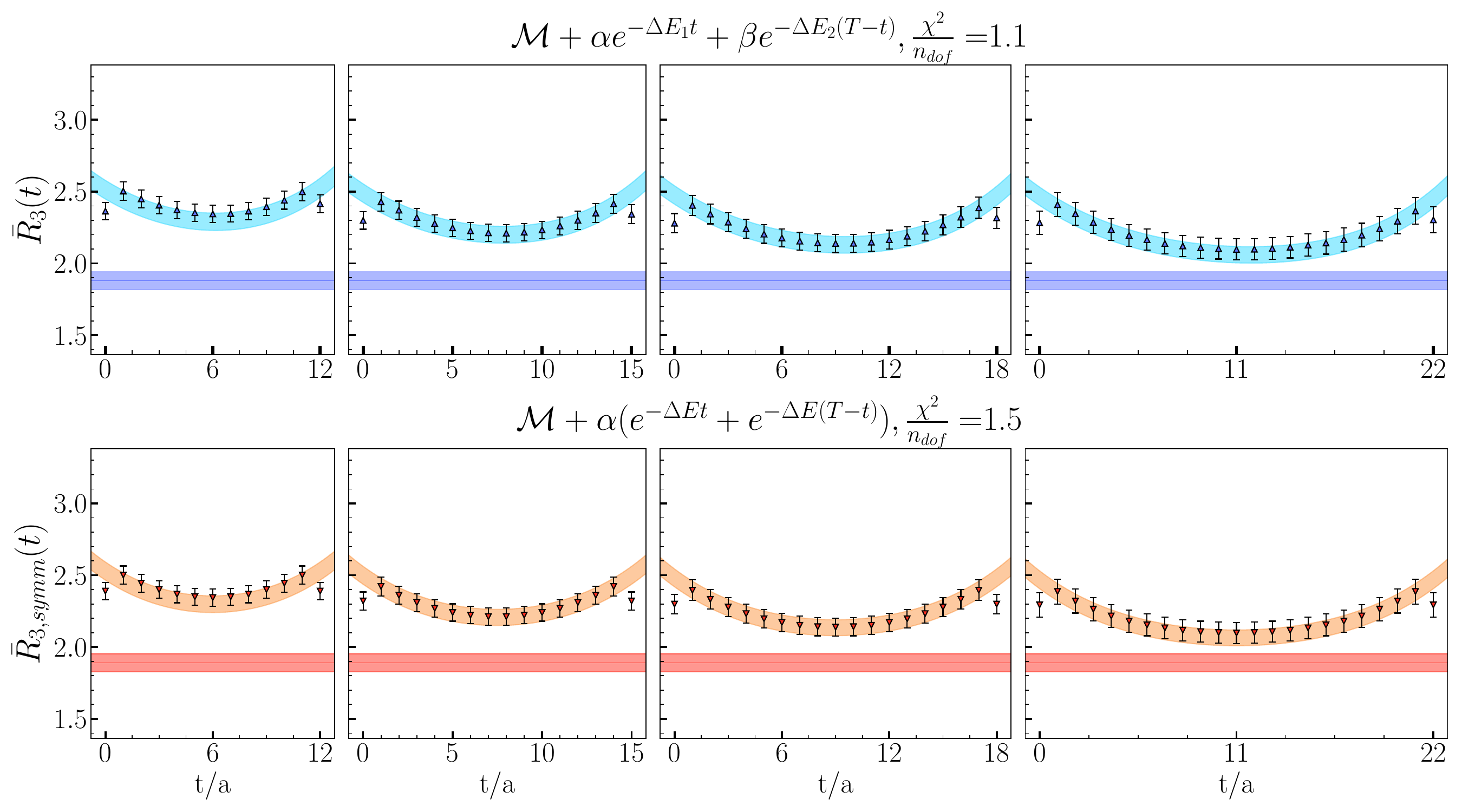}
    \caption{$B$ meson matrix element featuring the full renormalized EM current used to extract the charge form factor $F_C$ on the $\frac{L}{2\pi}\vec p_1 = (1,1,1)$ momentum shell. Other elements are the same as in Fig.~\ref{fig:ME3}.}
    \label{fig:ME5}
\end{figure*}

\begin{figure*}
    \includegraphics[width=0.93\textwidth]{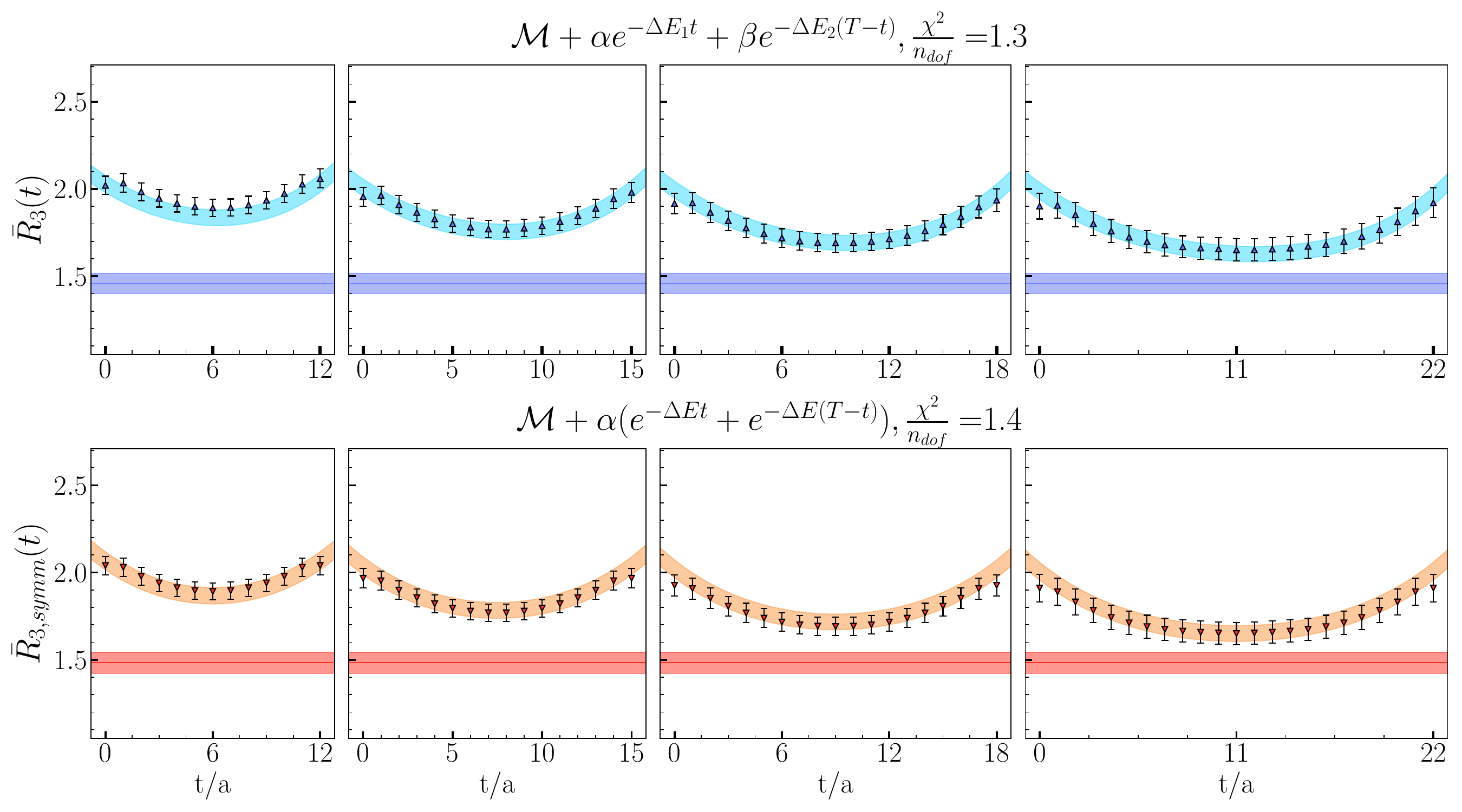}
    \caption{$B^*$ matrix element featuring the full renormalized EM current used to extract the first form factor $F_1$ on the $\frac{L}{2\pi}\vec p_1 = (1,1,1)$ momentum shell. Other elements are the same as in Fig.~\ref{fig:ME3}.}
    \label{fig:ME6}
\end{figure*}

\FloatBarrier

\supsection{Direction-averaging of three-point correlation function ratios}\label{sec:avgtables}
This section features Tables \ref{tab:tab1}-\ref{tab:tab3}, listing explicitly all relevant weights used in the direction-averaging of the three-point correlator ratios, as described in Section \ref{sec:Fitting three-point correlator data}.

\FloatBarrier

\begingroup
\setlength{\tabcolsep}{16pt}
\begin{longtable}{c l  l c l l c}
    \caption{List of all combinations of momenta $\vec p_2, \vec p_1$, irreps $\Lambda_2, \Lambda_1$ and rows $r_2, r_1$ at the sink and the source, respectively, that are relevant for the extraction of the $F_1$ form factor of the $T_{bb}$ from the ratios $R_3^\mu$. Before fitting the data, we averaged all ratios within each momentum shell that leads to the equivalent values of $Q^2$. Each shell is shown separated from others by double horizontal bars. The rightmost column, labeled by $w$, lists the weights that were used in averaging. The weights reflect the differences in polarizations and irrep projections that arise in each matrix element and correspond precisely to ones defined in eq.~\eqref{eq:avgratio}. Note that the relative weights are all normalized respective to the first entry in each momentum shell. To avoid excessive redundancy, momenta and irrep labels that are identical for all entries within a shell are listed only once.}\\
    \multicolumn{7}{c}{$\langle T_{bb} (p_2, \Lambda_2, r_2 \rangle) | \hat{\jmath}_{EM}^\mu | T_{bb} (p_1, \Lambda_1, r_1 \rangle), \ \mu \equiv 0$}  \\ 
    \toprule
    $\frac{L}{2\pi}\vec p_2$ & $\Lambda_2$ & $r_2$ & $\frac{L}{2\pi}\vec p_1$ & $\Lambda_1$ & $r_1$ & $w$ \\
    \toprule
    $(0, 0, 0)$ & $T_1^+$ & $1$ & $(0, 0, 0)$ & $T_1^+$ & $1$ & $1$ \\ 
    {} & {} & $2$ & {} & {} & $2$ & $1$ \\ 
    {} & {} & $3$ & {} & {} & $3$ & $1$ \\ 
    \midrule
    \midrule
    $(0, 0, 0)$ & $T_1^+$ & $1$ & $(0, 0, 1)$ & $E$ & $2$ & $1$ \\ 
    {} & {} & $1$ & $(0, 0, -1)$ & {} & $2$ & $1$ \\ 
    {} & {} & $1$ & $(0, 1, 0)$ & {} & $2$ & $1$ \\ 
    {} & {} & $1$ & $(0, -1, 0)$ & {} & $2$ & $1$ \\ 
    {} & {} & $2$ & $(0, 0, 1)$ & {} & $1$ & $1$ \\ 
    {} & {} & $2$ & $(0, 0, -1)$ & {} & $1$ & $1$ \\ 
    {} & {} & $2$ & $(1, 0, 0)$ & {} & $2$ & $1$ \\ 
    {} & {} & $2$ & $(-1, 0, 0)$ & {} & $2$ & $1$ \\ 
    {} & {} & $3$ & $(0, 1, 0)$ & {} & $1$ & $1$ \\ 
    {} & {} & $3$ & $(0, -1, 0)$ & {} & $1$ & $1$ \\ 
    {} & {} & $3$ & $(1, 0, 0)$ & {} & $1$ & $1$ \\ 
    {} & {} & $3$ & $(-1, 0, 0)$ & {} & $1$ & $1$ \\ 
    \midrule
    \midrule
    $(0, 0, 0)$ & $T_1^+$ & $1$ & $(0, 1, 1)$ & $B_1$ & $1$ & $1$ \\ 
    {} & {} & $1$ & $(0, 1, -1)$ & $B_1$ & $1$ & $1$ \\ 
    {} & {} & $1$ & $(0, -1, 1)$ & $B_1$ & $1$ & $1$ \\ 
    {} & {} & $1$ & $(0, -1, -1)$ & $B_1$ & $1$ & $1$ \\ 
    {} & {} & $1$ & $(1, 0, 1)$ & $B_2$ & $1$ & $\sqrt{2}$ \\ 
    {} & {} & $1$ & $(1, 0, -1)$ & $B_2$ & $1$ & $\sqrt{2}$ \\ 
    {} & {} & $1$ & $(-1, 0, 1)$ & $B_2$ & $1$ & $\sqrt{2}$ \\ 
    {} & {} & $1$ & $(-1, 0, -1)$ & $B_2$ & $1$ & $\sqrt{2}$ \\ 
    {} & {} & $1$ & $(1, 1, 0)$ & $B_2$ & $1$ & $\sqrt{2}$ \\ 
    {} & {} & $1$ & $(1, -1, 0)$ & $B_2$ & $1$ & $\sqrt{2}$ \\ 
    {} & {} & $1$ & $(-1, 1, 0)$ & $B_2$ & $1$ & $\sqrt{2}$ \\ 
    {} & {} & $1$ & $(-1, -1, 0)$ & $B_2$ & $1$ & $\sqrt{2}$ \\ 
    {} & {} & $2$ & $(0, 1, 1)$ & $B_2$ & $1$ & $\sqrt{2}$ \\ 
    {} & {} & $2$ & $(0, 1, -1)$ & $B_2$ & $1$ & $\sqrt{2}$ \\ 
    {} & {} & $2$ & $(0, -1, 1)$ & $B_2$ & $1$ & $\sqrt{2}$ \\ 
    {} & {} & $2$ & $(0, -1, -1)$ & $B_2$ & $1$ & $\sqrt{2}$ \\ 
    {} & {} & $2$ & $(1, 0, 1)$ & $B_1$ & $1$ & $1$ \\ 
    {} & {} & $2$ & $(1, 0, -1)$ & $B_1$ & $1$ & $1$ \\ 
    {} & {} & $2$ & $(-1, 0, 1)$ & $B_1$ & $1$ & $1$ \\ 
    {} & {} & $2$ & $(-1, 0, -1)$ & $B_1$ & $1$ & $1$ \\ 
    {} & {} & $2$ & $(1, 1, 0)$ & $B_2$ & $1$ & $-\sqrt{2}$ \\ 
    {} & {} & $2$ & $(1, -1, 0)$ & $B_2$ & $1$ & $\sqrt{2}$ \\ 
    {} & {} & $2$ & $(-1, 1, 0)$ & $B_2$ & $1$ & $\sqrt{2}$ \\ 
    {} & {} & $2$ & $(-1, -1, 0)$ & $B_2$ & $1$ & $-\sqrt{2}$ \\ 
    {} & {} & $3$ & $(0, 1, 1)$ & $B_2$ & $1$ & $-\sqrt{2}$ \\ 
    {} & {} & $3$ & $(0, 1, -1)$ & $B_2$ & $1$ & $\sqrt{2}$ \\ 
    {} & {} & $3$ & $(0, -1, 1)$ & $B_2$ & $1$ & $\sqrt{2}$ \\ 
    {} & {} & $3$ & $(0, -1, -1)$ & $B_2$ & $1$ & $-\sqrt{2}$ \\ 
    {} & {} & $3$ & $(1, 0, 1)$ & $B_2$ & $1$ & $-\sqrt{2}$ \\ 
    {} & {} & $3$ & $(1, 0, -1)$ & $B_2$ & $1$ & $\sqrt{2}$ \\ 
    {} & {} & $3$ & $(-1, 0, 1)$ & $B_2$ & $1$ & $\sqrt{2}$ \\ 
    {} & {} & $3$ & $(-1, 0, -1)$ & $B_2$ & $1$ & $-\sqrt{2}$ \\ 
    {} & {} & $3$ & $(1, 1, 0)$ & $B_1$ & $1$ & $1$ \\ 
    {} & {} & $3$ & $(1, -1, 0)$ & $B_1$ & $1$ & $1$ \\ 
    {} & {} & $3$ & $(-1, 1, 0)$ & $B_1$ & $1$ & $1$ \\ 
    {} & {} & $3$ & $(-1, -1, 0)$ & $B_1$ & $1$ & $1$ \\ 
    \midrule
    \midrule
    $(0, 0, 0)$ & $T_1^+$ & $1$ & $(1, 1, 1)$ & $E$ & $1$ & $1$ \\ 
    {} & {} & $1$ & $(1, 1, -1)$ & {} & $1$ & $1$ \\ 
    {} & {} & $1$ & $(1, -1, 1)$ & {} & $1$ & $1$ \\ 
    {} & {} & $1$ & $(1, -1, -1)$ & {} & $1$ & $1$ \\ 
    {} & {} & $1$ & $(-1, 1, 1)$ & {} & $1$ & $1$ \\ 
    {} & {} & $1$ & $(-1, 1, -1)$ & {} & $1$ & $1$ \\ 
    {} & {} & $1$ & $(-1, -1, 1)$ & {} & $1$ & $1$ \\ 
    {} & {} & $1$ & $(-1, -1, -1)$ & {} & $1$ & $1$ \\ 
    {} & {} & $2$ & $(1, 1, 1)$ & {} & $2$ & $\frac{2}{\sqrt{3}}$ \\ 
    {} & {} & $2$ & $(1, 1, 1)$ & {} & $1$ & $-2$ \\ 
    {} & {} & $2$ & $(1, 1, -1)$ & {} & $2$ & $\frac{2}{\sqrt{3}}$ \\ 
    {} & {} & $2$ & $(1, 1, -1)$ & {} & $1$ & $-2$ \\ 
    {} & {} & $2$ & $(1, -1, 1)$ & {} & $1$ & $2$ \\ 
    {} & {} & $2$ & $(1, -1, 1)$ & {} & $2$ & $\frac{2}{\sqrt{3}}$ \\ 
    {} & {} & $2$ & $(1, -1, -1)$ & {} & $1$ & $2$ \\ 
    {} & {} & $2$ & $(1, -1, -1)$ & {} & $2$ & $\frac{2}{\sqrt{3}}$ \\ 
    {} & {} & $2$ & $(-1, 1, 1)$ & {} & $1$ & $2$ \\ 
    {} & {} & $2$ & $(-1, 1, 1)$ & {} & $2$ & $\frac{2}{\sqrt{3}}$ \\ 
    {} & {} & $2$ & $(-1, 1, -1)$ & {} & $1$ & $2$ \\ 
    {} & {} & $2$ & $(-1, 1, -1)$ & {} & $2$ & $\frac{2}{\sqrt{3}}$ \\ 
    {} & {} & $2$ & $(-1, -1, 1)$ & {} & $2$ & $\frac{2}{\sqrt{3}}$ \\ 
    {} & {} & $2$ & $(-1, -1, 1)$ & {} & $1$ & $-2$ \\ 
    {} & {} & $2$ & $(-1, -1, -1)$ & {} & $2$ & $\frac{2}{\sqrt{3}}$ \\ 
    {} & {} & $2$ & $(-1, -1, -1)$ & {} & $1$ & $-2$ \\ 
    {} & {} & $3$ & $(1, 1, 1)$ & {} & $2$ & $-\frac{2}{\sqrt{3}}$ \\ 
    {} & {} & $3$ & $(1, 1, 1)$ & {} & $1$ & $-2$ \\ 
    {} & {} & $3$ & $(1, 1, -1)$ & {} & $2$ & $\frac{2}{\sqrt{3}}$ \\ 
    {} & {} & $3$ & $(1, 1, -1)$ & {} & $1$ & $2$ \\ 
    {} & {} & $3$ & $(1, -1, 1)$ & {} & $1$ & $-2$ \\ 
    {} & {} & $3$ & $(1, -1, 1)$ & {} & $2$ & $\frac{2}{\sqrt{3}}$ \\ 
    {} & {} & $3$ & $(1, -1, -1)$ & {} & $1$ & $2$ \\ 
    {} & {} & $3$ & $(1, -1, -1)$ & {} & $2$ & $-\frac{2}{\sqrt{3}}$ \\ 
    {} & {} & $3$ & $(-1, 1, 1)$ & {} & $1$ & $2$ \\ 
    {} & {} & $3$ & $(-1, 1, 1)$ & {} & $2$ & $-\frac{2}{\sqrt{3}}$ \\ 
    {} & {} & $3$ & $(-1, 1, -1)$ & {} & $1$ & $-2$ \\ 
    {} & {} & $3$ & $(-1, 1, -1)$ & {} & $2$ & $\frac{2}{\sqrt{3}}$ \\ 
    {} & {} & $3$ & $(-1, -1, 1)$ & {} & $2$ & $\frac{2}{\sqrt{3}}$ \\ 
    {} & {} & $3$ & $(-1, -1, 1)$ & {} & $1$ & $2$ \\ 
    {} & {} & $3$ & $(-1, -1, -1)$ & {} & $2$ & $-\frac{2}{\sqrt{3}}$ \\ 
    {} & {} & $3$ & $(-1, -1, -1)$ & {} & $1$ & $-2$ \\ 
    \midrule
    \midrule
    $(0, 0, 0)$ & $T_1^+$ & $1$ & $(0, 0, 2)$ & $E$ & $2$ & $1$ \\ 
    {} & {} & $1$ & $(0, 0, -2)$ & {} & $2$ & $1$ \\ 
    {} & {} & $1$ & $(0, 2, 0)$ & {} & $2$ & $1$ \\ 
    {} & {} & $1$ & $(0, -2, 0)$ & {} & $2$ & $1$ \\ 
    {} & {} & $2$ & $(0, 0, 2)$ & {} & $1$ & $1$ \\ 
    {} & {} & $2$ & $(0, 0, -2)$ & {} & $1$ & $1$ \\ 
    {} & {} & $2$ & $(2, 0, 0)$ & {} & $2$ & $1$ \\ 
    {} & {} & $2$ & $(-2, 0, 0)$ & {} & $2$ & $1$ \\ 
    {} & {} & $3$ & $(0, 2, 0)$ & {} & $1$ & $1$ \\ 
    {} & {} & $3$ & $(0, -2, 0)$ & {} & $1$ & $1$ \\ 
    {} & {} & $3$ & $(2, 0, 0)$ & {} & $1$ & $1$ \\ 
    {} & {} & $3$ & $(-2, 0, 0)$ & {} & $1$ & $1$ \\ 
    \midrule
    \midrule
    $(0, 0, 0)$ & $T_1^+$ & $1$ & $(0, 0, 3)$ & $E$ & $2$ & $1$ \\ 
    {} & {} & $1$ & $(0, 0, -3)$ & {} & $2$ & $1$ \\ 
    {} & {} & $1$ & $(0, 3, 0)$ & {} & $2$ & $1$ \\ 
    {} & {} & $1$ & $(0, -3, 0)$ & {} & $2$ & $1$ \\ 
    {} & {} & $2$ & $(0, 0, 3)$ & {} & $1$ & $1$ \\ 
    {} & {} & $2$ & $(0, 0, -3)$ & {} & $1$ & $1$ \\ 
    {} & {} & $2$ & $(3, 0, 0)$ & {} & $2$ & $1$ \\ 
    {} & {} & $2$ & $(-3, 0, 0)$ & {} & $2$ & $1$ \\ 
    {} & {} & $3$ & $(0, 3, 0)$ & {} & $1$ & $1$ \\ 
    {} & {} & $3$ & $(0, -3, 0)$ & {} & $1$ & $1$ \\ 
    {} & {} & $3$ & $(3, 0, 0)$ & {} & $1$ & $1$ \\ 
    {} & {} & $3$ & $(-3, 0, 0)$ & {} & $1$ & $1$ \\ 
    \bottomrule
    \label{tab:tab1}
\end{longtable}
\endgroup

\begingroup
\setlength{\tabcolsep}{16pt}
\begin{longtable}{c l  l c l l c}
    \caption{List of all combinations of momenta $\vec p_2, \vec p_1$, irreps $\Lambda_2, \Lambda_1$ and rows $r_2, r_1$ at the sink and the source, respectively, that are relevant for the extraction of the $F_2$ form factor of the $T_{bb}$ from the ratios $R_3^\mu$. The elements of the table are organized in the same way as in Table \ref{tab:tab1}.}\\
    \multicolumn{7}{c}{$\langle T_{bb} (p_2, \Lambda_2, r_2 \rangle) | \hat{\jmath}_{EM}^\mu | T_{bb} (p_1, \Lambda_1, r_1 \rangle), \ \mu \equiv 2$}  \\ 
    \toprule
    $\frac{L}{2\pi}\vec p_2$ & $\Lambda_2$ & $r_2$ & $\frac{L}{2\pi}\vec p_1$ & $\Lambda_1$ & $r_1$ & $w$ \\
    \toprule
    $(0, 0, 0)$ & $T_1^+$ & $1$ & $(1, 0, 0)$ & $E$ & $2$ & $1$ \\ 
    {} & {} & $1$ & $(-1, 0, 0)$ & $E$ & $2$ & $-1$ \\ 
    {} & {} & $2$ & $(0, 0, 1)$ & $A_2$ & $1$ & $-1$ \\ 
    {} & {} & $2$ & $(0, 0, -1)$ & $A_2$ & $1$ & $1$ \\ 
    {} & {} & $2$ & $(1, 0, 0)$ & $A_2$ & $1$ & $-1$ \\ 
    {} & {} & $2$ & $(-1, 0, 0)$ & $A_2$ & $1$ & $1$ \\ 
    {} & {} & $3$ & $(0, 0, 1)$ & $E$ & $1$ & $1$ \\ 
    {} & {} & $3$ & $(0, 0, -1)$ & $E$ & $1$ & $-1$ \\ 
    \midrule
    \midrule
    $(0, 0, 0)$ & $T_1^+$ & $1$ & $(1, 0, 1)$ & $B_1$ & $1$ & $1$ \\ 
    {} & {} & $1$ & $(1, 0, -1)$ & $B_1$ & $1$ & $1$ \\ 
    {} & {} & $1$ & $(-1, 0, 1)$ & $B_1$ & $1$ & $-1$ \\ 
    {} & {} & $1$ & $(-1, 0, -1)$ & $B_1$ & $1$ & $-1$ \\ 
    {} & {} & $2$ & $(1, 0, 1)$ & $A_2$ & $1$ & $-\frac{1}{\sqrt{2}}$ \\ 
    {} & {} & $2$ & $(1, 0, -1)$ & $A_2$ & $1$ & $-\frac{1}{\sqrt{2}}$ \\ 
    {} & {} & $2$ & $(-1, 0, 1)$ & $A_2$ & $1$ & $\frac{1}{\sqrt{2}}$ \\ 
    {} & {} & $2$ & $(-1, 0, -1)$ & $A_2$ & $1$ & $\frac{1}{\sqrt{2}}$ \\ 
    {} & {} & $3$ & $(1, 0, 1)$ & $B_1$ & $1$ & $1$ \\ 
    {} & {} & $3$ & $(1, 0, -1)$ & $B_1$ & $1$ & $-1$ \\ 
    {} & {} & $3$ & $(-1, 0, 1)$ & $B_1$ & $1$ & $1$ \\ 
    {} & {} & $3$ & $(-1, 0, -1)$ & $B_1$ & $1$ & $-1$ \\ 
    \midrule
    \midrule
    $(0, 0, 0)$ & $T_1^+$ & $1$ & $(1, 1, 1)$ & $E$ & $2$ & $1$ \\ 
    {} & {} & $1$ & $(1, 1, -1)$ & {} & $2$ & $1$ \\ 
    {} & {} & $1$ & $(1, -1, 1)$ & {} & $2$ & $1$ \\ 
    {} & {} & $1$ & $(1, -1, -1)$ & {} & $2$ & $1$ \\ 
    {} & {} & $1$ & $(-1, 1, 1)$ & {} & $2$ & $-1$ \\ 
    {} & {} & $1$ & $(-1, 1, -1)$ & {} & $2$ & $-1$ \\ 
    {} & {} & $1$ & $(-1, -1, 1)$ & {} & $2$ & $-1$ \\ 
    {} & {} & $1$ & $(-1, -1, -1)$ & {} & $2$ & $-1$ \\ 
    \midrule
    \midrule
    $(0, 0, 0)$ & $T_1^+$ & $1$ & $(2, 0, 0)$ & $E$ & $2$ & $1$ \\ 
    {} & {} & $1$ & $(-2, 0, 0)$ & $E$ & $2$ & $-1$ \\ 
    {} & {} & $2$ & $(0, 0, 2)$ & $A_2$ & $1$ & $-1$ \\ 
    {} & {} & $2$ & $(0, 0, -2)$ & $A_2$ & $1$ & $1$ \\ 
    {} & {} & $2$ & $(2, 0, 0)$ & $A_2$ & $1$ & $-1$ \\ 
    {} & {} & $2$ & $(-2, 0, 0)$ & $A_2$ & $1$ & $1$ \\ 
    {} & {} & $3$ & $(0, 0, 2)$ & $E$ & $1$ & $1$ \\ 
    {} & {} & $3$ & $(0, 0, -2)$ & $E$ & $1$ & $-1$ \\ 
    \midrule
    \midrule
    $(0, 0, 0)$ & $T_1^+$ & $1$ & $(3, 0, 0)$ & $E$ & $2$ & $1$ \\ 
    {} & {} & $1$ & $(-3, 0, 0)$ & $E$ & $2$ & $-1$ \\ 
    {} & {} & $2$ & $(0, 0, 3)$ & $A_2$ & $1$ & $-1$ \\ 
    {} & {} & $2$ & $(0, 0, -3)$ & $A_2$ & $1$ & $1$ \\ 
    {} & {} & $2$ & $(3, 0, 0)$ & $A_2$ & $1$ & $-1$ \\ 
    {} & {} & $2$ & $(-3, 0, 0)$ & $A_2$ & $1$ & $1$ \\ 
    {} & {} & $3$ & $(0, 0, 3)$ & $E$ & $1$ & $1$ \\ 
    {} & {} & $3$ & $(0, 0, -3)$ & $E$ & $1$ & $-1$ \\ 
    \bottomrule
    \label{tab:tab2}
\end{longtable}
\endgroup

\begingroup
\setlength{\tabcolsep}{16pt}
\begin{longtable}{c l  l c l l c}
    \caption{List of all combinations of momenta $\vec p_2, \vec p_1$, irreps $\Lambda_2, \Lambda_1$ and rows $r_2, r_1$ at the sink and the source, respectively, that are relevant for the extraction of the $F_3$ form factor of the $T_{bb}$ from the ratios $R_3^\mu$. The elements of the table are organized in the same way as in Table \ref{tab:tab1}.}\\
    \multicolumn{7}{c}{$\langle T_{bb} (p_2, \Lambda_2, r_2 \rangle) | \hat{\jmath}_{EM}^\mu | T_{bb} (p_1, \Lambda_1, r_1 \rangle), \ \mu \equiv 0$}  \\ 
    \toprule
    $\frac{L}{2\pi}\vec p_2$ & $\Lambda_2$ & $r_2$ & $\frac{L}{2\pi}\vec p_1$ & $\Lambda_1$ & $r_1$ & $w$ \\
    \toprule
    $(0, 0, 0)$ & $T_1^+$ & $1$ & $(1, 0, 0)$ & $A_2$ & $1$ & $1$ \\ 
    {} & {} & $1$ & $(-1, 0, 0)$ & {} & $1$ & $1$ \\ 
    {} & {} & $2$ & $(0, 1, 0)$ & {} & $1$ & $1$ \\ 
    {} & {} & $2$ & $(0, -1, 0)$ & {} & $1$ & $1$ \\ 
    {} & {} & $3$ & $(0, 0, 1)$ & {} & $1$ & $1$ \\ 
    {} & {} & $3$ & $(0, 0, -1)$ & {} & $1$ & $1$ \\ 
    \midrule
    \midrule
    $(0, 0, 0)$ & $T_1^+$ & $1$ & $(1, 0, 1)$ & $A_2$ & $1$ & $1$ \\ 
    {} & {} & $1$ & $(1, 0, -1)$ & {} & $1$ & $1$ \\ 
    {} & {} & $1$ & $(-1, 0, 1)$ & {} & $1$ & $1$ \\ 
    {} & {} & $1$ & $(-1, 0, -1)$ & {} & $1$ & $1$ \\ 
    {} & {} & $1$ & $(1, 1, 0)$ & {} & $1$ & $1$ \\ 
    {} & {} & $1$ & $(1, -1, 0)$ & {} & $1$ & $1$ \\ 
    {} & {} & $1$ & $(-1, 1, 0)$ & {} & $1$ & $1$ \\ 
    {} & {} & $1$ & $(-1, -1, 0)$ & {} & $1$ & $1$ \\ 
    {} & {} & $2$ & $(0, 1, 1)$ & {} & $1$ & $1$ \\ 
    {} & {} & $2$ & $(0, 1, -1)$ & {} & $1$ & $1$ \\ 
    {} & {} & $2$ & $(0, -1, 1)$ & {} & $1$ & $1$ \\ 
    {} & {} & $2$ & $(0, -1, -1)$ & {} & $1$ & $1$ \\ 
    {} & {} & $2$ & $(1, 1, 0)$ & {} & $1$ & $1$ \\ 
    {} & {} & $2$ & $(1, -1, 0)$ & {} & $1$ & $-1$ \\ 
    {} & {} & $2$ & $(-1, 1, 0)$ & {} & $1$ & $-1$ \\ 
    {} & {} & $2$ & $(-1, -1, 0)$ & {} & $1$ & $1$ \\ 
    {} & {} & $3$ & $(0, 1, 1)$ & {} & $1$ & $1$ \\ 
    {} & {} & $3$ & $(0, 1, -1)$ & {} & $1$ & $-1$ \\ 
    {} & {} & $3$ & $(0, -1, 1)$ & {} & $1$ & $-1$ \\ 
    {} & {} & $3$ & $(0, -1, -1)$ & {} & $1$ & $1$ \\ 
    {} & {} & $3$ & $(1, 0, 1)$ & {} & $1$ & $1$ \\ 
    {} & {} & $3$ & $(1, 0, -1)$ & {} & $1$ & $-1$ \\ 
    {} & {} & $3$ & $(-1, 0, 1)$ & {} & $1$ & $-1$ \\ 
    {} & {} & $3$ & $(-1, 0, -1)$ & {} & $1$ & $1$ \\ 
    \midrule
    \midrule
    $(0, 0, 0)$ & $T_1^+$ & $1$ & $(1, 1, 1)$ & $A_2$ & $1$ & $1$ \\ 
    {} & {} & $1$ & $(1, 1, -1)$ & {} & $1$ & $1$ \\ 
    {} & {} & $1$ & $(1, -1, 1)$ & {} & $1$ & $1$ \\ 
    {} & {} & $1$ & $(1, -1, -1)$ & {} & $1$ & $1$ \\ 
    {} & {} & $1$ & $(-1, 1, 1)$ & {} & $1$ & $1$ \\ 
    {} & {} & $1$ & $(-1, 1, -1)$ & {} & $1$ & $1$ \\ 
    {} & {} & $1$ & $(-1, -1, 1)$ & {} & $1$ & $1$ \\ 
    {} & {} & $1$ & $(-1, -1, -1)$ & {} & $1$ & $1$ \\ 
    {} & {} & $2$ & $(1, 1, 1)$ & {} & $1$ & $1$ \\ 
    {} & {} & $2$ & $(1, 1, -1)$ & {} & $1$ & $1$ \\ 
    {} & {} & $2$ & $(1, -1, 1)$ & {} & $1$ & $-1$ \\ 
    {} & {} & $2$ & $(1, -1, -1)$ & {} & $1$ & $-1$ \\ 
    {} & {} & $2$ & $(-1, 1, 1)$ & {} & $1$ & $-1$ \\ 
    {} & {} & $2$ & $(-1, 1, -1)$ & {} & $1$ & $-1$ \\ 
    {} & {} & $2$ & $(-1, -1, 1)$ & {} & $1$ & $1$ \\ 
    {} & {} & $2$ & $(-1, -1, -1)$ & {} & $1$ & $1$ \\ 
    {} & {} & $3$ & $(1, 1, 1)$ & {} & $1$ & $1$ \\ 
    {} & {} & $3$ & $(1, 1, -1)$ & {} & $1$ & $-1$ \\ 
    {} & {} & $3$ & $(1, -1, 1)$ & {} & $1$ & $1$ \\ 
    {} & {} & $3$ & $(1, -1, -1)$ & {} & $1$ & $-1$ \\ 
    {} & {} & $3$ & $(-1, 1, 1)$ & {} & $1$ & $-1$ \\ 
    {} & {} & $3$ & $(-1, 1, -1)$ & {} & $1$ & $1$ \\ 
    {} & {} & $3$ & $(-1, -1, 1)$ & {} & $1$ & $-1$ \\ 
    {} & {} & $3$ & $(-1, -1, -1)$ & {} & $1$ & $1$ \\ 
    \midrule
    \midrule
    $(0, 0, 0)$ & $T_1^+$ & $1$ & $(2, 0, 0)$ & $A_2$ & $1$ & $1$ \\ 
    {} & {} & $1$ & $(-2, 0, 0)$ & {} & $1$ & $1$ \\ 
    {} & {} & $2$ & $(0, 2, 0)$ & {} & $1$ & $1$ \\ 
    {} & {} & $2$ & $(0, -2, 0)$ & {} & $1$ & $1$ \\ 
    {} & {} & $3$ & $(0, 0, 2)$ & {} & $1$ & $1$ \\ 
    {} & {} & $3$ & $(0, 0, -2)$ & {} & $1$ & $1$ \\ 
    \midrule
    \midrule
    $(0, 0, 0)$ & $T_1^+$ & $1$ & $(3, 0, 0)$ & $A_2$ & $1$ & $1$ \\ 
    {} & {} & $1$ & $(-3, 0, 0)$ & {} & $1$ & $1$ \\ 
    {} & {} & $2$ & $(0, 3, 0)$ & {} & $1$ & $1$ \\ 
    {} & {} & $2$ & $(0, -3, 0)$ & {} & $1$ & $1$ \\ 
    {} & {} & $3$ & $(0, 0, 3)$ & {} & $1$ & $1$ \\ 
    {} & {} & $3$ & $(0, 0, -3)$ & {} & $1$ & $1$ \\ 
    \bottomrule
    \label{tab:tab3}
\end{longtable}
\endgroup

\end{document}